\documentclass[aps, prd, superscriptaddress, preprintnumbers, floatfix, longbibliography,10pt,twocolumn]{revtex4-2}

\usepackage{ graphicx}
\usepackage{  amsfonts,amsmath,amssymb,amstext}
\usepackage{  float,wrapfig}
\usepackage{  subfigure, psfrag}
\usepackage{  dsfont}
\usepackage{  color}
\usepackage{  comment}
\usepackage{  multirow}

\usepackage[colorlinks=true,urlcolor=blue,citecolor=blue,linkcolor=blue]{hyperref}
\usepackage{  bm}
\usepackage{  mathtools}

\usepackage[colorlinks=true,urlcolor=blue,citecolor=blue,linkcolor=blue]{hyperref}
\hypersetup{breaklinks=true}

\definecolor{purple}{rgb}{0.8,0,0.6}
\definecolor{darkgreen}{rgb}{0.00,0.6,0.00}

\begin{document}

\title{Correlated phases in rhombohedral multilayer graphene}

\author{Arsen~Herasymchuk}
\email{arsen.herasymchuk@epfl.ch}
\affiliation{Bogolyubov Institute for Theoretical Physics, Kyiv, 03143, Ukraine}

\author{Sergei~G.~Sharapov}
\email{sharapov@bitp.kyiv.ua}
\affiliation{Bogolyubov Institute for Theoretical Physics, Kyiv, 03143, Ukraine}
\affiliation{
Kyiv Academic University, 03142 Kyiv, Ukraine}

\author{Oleg~V.~Yazyev}
\email{oleg.yazyev@epfl.ch}
\affiliation{Institute of Physics, \'{E}cole Polytechnique F\'{e}d\'{e}rale de Lausanne (EPFL), CH-1015 Lausanne, Switzerland}

\author{Yaroslav~Zhumagulov}
\email{iaroslav.zhumagulov@epfl.ch}
\affiliation{Institute of Physics, \'{E}cole Polytechnique F\'{e}d\'{e}rale de Lausanne (EPFL), CH-1015 Lausanne, Switzerland}

\begin{abstract}
We investigate the emergence of correlated electron phases in rhombohedral \(N\)-layer graphene due to two-valley Coulomb interactions within a low-energy \(k \cdot p\) framework. Analytical expressions for Lindhard susceptibilities in intra- and intervalley channels are derived, and the critical temperatures for phase transitions are estimated using both the random phase approximation (RPA) and the parquet approximation (PA). Within RPA, only Stoner and intervalley coherent (IVC) phases are supported, while the PA reveals a richer phase structure including particle-particle (PP) channel instabilities. We establish a general scaling law for the critical temperature with respect to layer number \(N\), highlighting an upper bound as \(N \rightarrow \infty\), and demonstrate a non-monotonic decrease of the critical temperature with increasing chemical potential. The PA uncovers the role of interaction symmetry: \(SU(4)\)-symmetric interactions favor intervalley Stoner order in the density channel, whereas \(SU(2) \times SU(2)\)-symmetric interactions permit a broader set of phases. A crossover in the dominant instability occurs in the particle-hole channel at a critical layer number, suggesting the emergence of magnetic or IVC phases in thicker systems. We also identify conditions under which pair-density wave (PDW) order could form in the PP channel, though its physical realization may be constrained.
\end{abstract}
\date{\today }\maketitle

\section{Introduction}
Rhombohedral multilayer graphene (R$N$G) has recently emerged as a versatile platform for realizing strongly correlated electronic phases in a structurally simple crystalline system ~\cite{Zhou2021_1, Zhou2021_2, Pierucci2015, Shi2020}, free from the moiré superlattices that define twisted graphene~\cite{Dean2013, Kim2016, Cao2018_1, Cao2018_2}. In the rhombohedral stacking sequence, the low-energy electronic states near the $K$ and $K'$ valleys are hosted in nearly flat conduction and valence bands whose bandwidth decreases sharply with increasing $N$~\cite{Li2009, PhysRevB.82.035409, acs.nanolett.8b02530, Pierucci2015}. This band flattening enhances the density of states (DOS) near the charge neutrality point, giving rise to Van Hove singularities (VHS) and magnifying the role of electron–electron interactions.

Recent experiments on bilayer~\cite{Zhou2022, Zhang2023, Seiler2022}, trilayer~\cite{Zhou2021_1, Zhou2021_2}, quadralayer~\cite{Cao2023, Liu2023}, and pentalayer~\cite{Cao2023, Han2023} rhombohedral graphene have shown a range of interaction-driven phases, including half-metallic and quarter-metallic states~\cite{Zhou2021_1}, spin-polarized superconductivity~\cite{Zhou2022, Zhang2023, Seiler2022, Zhou2021_2, Han2025ChiralSuperconductivity, Yang2025superconductivity}, quantized anomalous Hall effects~\cite{science.adk9749, Winterer2024, Han2023}, magnetic and isospin order~\cite{Zhou2024ferromagnetism, Han2023multiferroicity, Liu2023, Seiler_2025}, and fractional Chern insulator~\cite{Xie2025}. The variety of these phases suggests that strong Coulomb interactions in R$N$G can stabilize competing broken-symmetry ground states, whose nature depends on the level of doping, displacement field, and layer number. The experimental tunability of these parameters makes R$N$G an ideal system for systematically probing electron correlation physics.

On the theoretical side, the interaction-driven instabilities in R$N$G have been studied using mean-field theory~\cite{christos2025finitemomentumpairingsuperlatticesuperconductivity, Dong2023_1, Dong2023_2, Huang2023,wy3f-hgr9, PhysRevB.105.L201107, Mu2025, PhysRevB.111.174523, Mayrhofer2025, PhysRevB.111.L081111, Kopnin2011}, Hartree–Fock calculations~\cite{PhysRevB.104.035404, Huang2025, PhysRevLett.133.206502}, renormalization group~\cite{PhysRevB.105.L081407, PhysRevLett.130.146001, PhysRevB.106.155115} and RPA~\cite{PhysRevB.106.155115, PhysRevLett.132.186401, PhysRevB.110.045427} approaches. These works have identified possible Stoner magnetism, superconductivity with unconventional pairing symmetries, and intervalley-coherent (IVC) phases as leading candidates for the ground state~\cite{Dong2023_2,PhysRevB.110.104420,PhysRevB.110.235433,PhysRevB.108.134503}. A unifying feature of these phases is their connection to the VHS structure, which is tunable by electrostatic gating and external fields. Compared to magic-angle twisted bilayer graphene (MATBG), where flat bands arise from moiré interference which also result in the emergence of correlated phases~\cite{PhysRevB.101.224513,PhysRevLett.128.227601}, R$N$G offers a simpler and more reproducible route to high-DOS regimes, while avoiding the fabrication challenges of twist-angle control and minimizing twist disorder~\cite{Beechem2014, Uri2020, Gadelha2021, PhysRevResearch.2.023325, Kazmierczak2021, PhysRevB.105.245408, PhysRevB.102.064501,PhysRevB.107.L081403}.

Despite recent advancements, the relationship between the critical temperature of correlated phases and the number of layers \( N \), as well as the behavior of phase diagrams with varying \( N \), especially in the limit as \( N \) approaches infinity, remains less explored~\cite{PhysRevB.108.144504,PhysRevB.110.L241401}. The limit of \( N \to \infty \) is important for identifying the scaling laws that govern correlated states across different layer thicknesses and for showing connections between few-layer graphene and bulk rhombohedral graphite. Additionally, understanding the interplay between competing instabilities in this limit requires a framework that can simultaneously address particle-hole and particle-particle interactions.

In this work, we investigate correlated phase formation in R$N$G driven by two-valley Coulomb interactions~\cite{PhysRevB.106.155115, PhysRevLett.132.186401, PhysRevB.110.045427, PhysRevB.110.L201113} within a low-energy $k\cdot p$ framework valid for $N \ge 3$~\cite{McCann_2013, PhysRevB.82.035409, PhysRevB.77.045429}. We derive analytical expressions for the Lindhard susceptibilities in both intra- and intervalley channels, enabling a systematic comparison between instabilities in the particle–hole (PH) and particle–particle (PP) sectors. Two complementary approximations are employed: the random phase approximation (RPA), which captures screening effects but restricts the competition between channels, and the parquet approximation (PA)~\cite{RevModPhys.90.025003, PhysRevB.86.125114, TheorMethodsSCEbook}, which incorporates the interplay between PH and PP processes.

Our analysis reveals several key results. Within RPA, only the Stoner and IVC phases are supported. The PA uncovers a richer phase structure, including PP instabilities such as superconductivity and possible pair-density wave (PDW) order. We derive a general scaling law for the critical temperature $T_c$ with respect to $N$, identifying an upper bound as $N \to \infty$. Furthermore, we find a non-monotonic dependence of $T_c$ on the chemical potential and show that the symmetry of the Coulomb interaction plays a decisive role: $SU(4)$-symmetric interactions favor intervalley Stoner order, whereas $SU(2)\times SU(2)$ symmetry admits a broader range of competing phases. Our results also predict a crossover in the dominant PH instability at a critical $N$, indicating a shift toward magnetic or IVC phases in thicker samples.

The paper is organized as follows. We describe the effective two-band low-energy model of R$N$G in Sec.~\ref{sec:Model}. In Sec.~\ref{sec:susceptibility}, we remind general properties of two-particle correlation functions and bare particle-hole and particle-particle susceptibilities are calculated. The random phase approximation is explored in Sec.~\ref{sec:rpa}. In Sec.~\ref{sec:parquet}, we study the correlated phase emergence by considering the parquet approximation. The results are summarized in Sec.~\ref{sec:summary}. Technical details are presented in Appendixes~\ref{app:Kubo} and \ref{app:explicit-vertices}. Throughout this paper, we use $\hbar=k_{\rm B}=1$.

\section{Effective two-band Hamiltonian of rhombohedral $N$-layer graphene}
\label{sec:Model}

\begin{figure}[t]
\includegraphics[width=.45\textwidth]{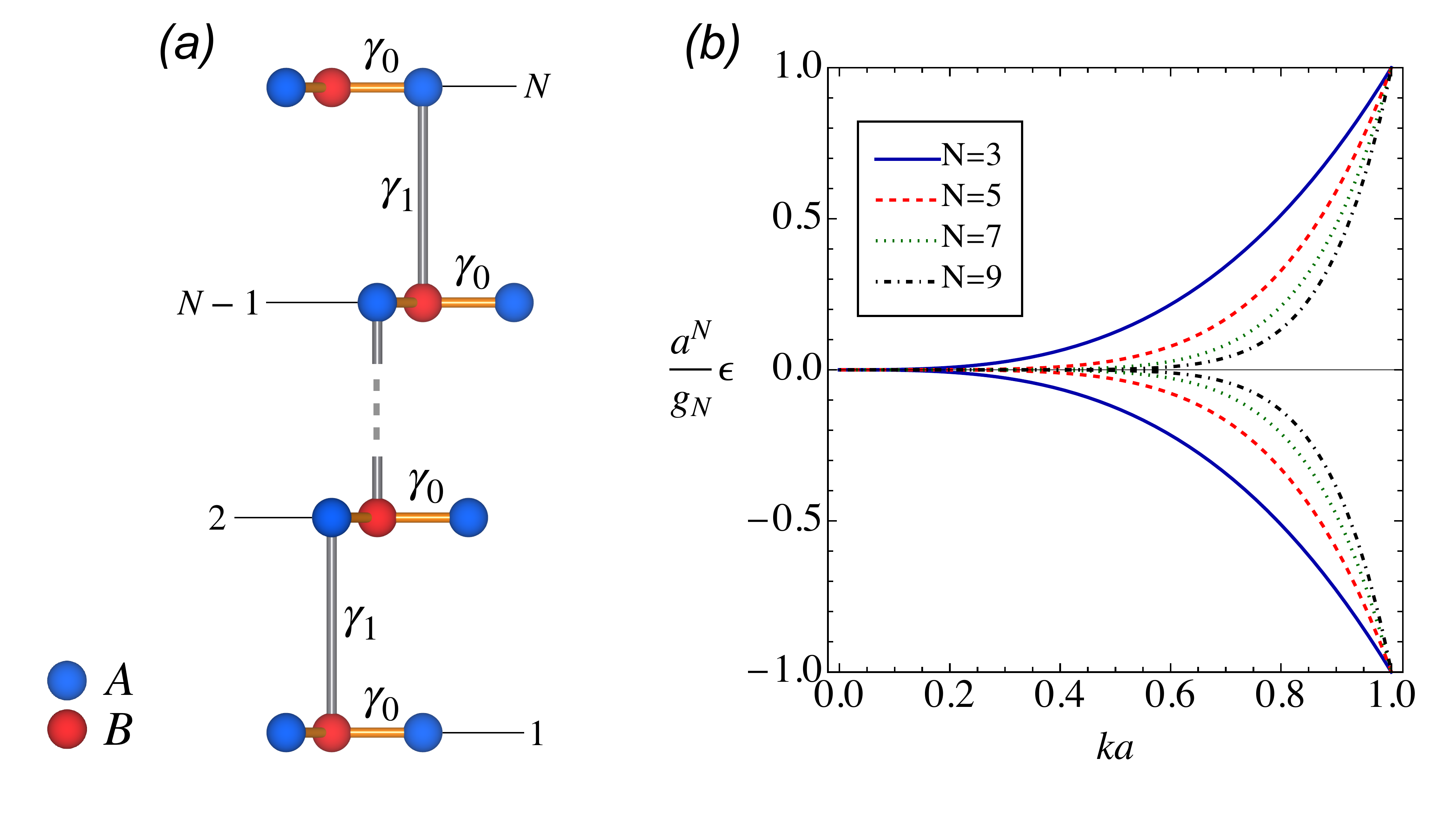}
\caption{(a) The unit cell of rhombohedral $N$-layer graphene that corresponds to Hamiltonian density given by Eq.~(\ref{eq:Nlayer-graphene-model}) with the nearest-neighboring intralayer $\gamma_0$ and interlayer $\gamma_1$ orbital hoppings. (b) The dispersion relation of the effective low-energy model given by Eq.~(\ref{eq:dispresion}) for four values of $N$. Here, $a$ is the lattice constant, the energy $\varepsilon$ is in the units of $g_N/a^N$.}
\label{fig:band-structure}
\end{figure}

We consider rhombohedral (ABC-stacked) $N$-layer graphene ($N \ge 3$). 
In each graphene layer, the unit cell contains two carbon
atoms $A$ and $B$ that are not equivalent, as shown in Fig.~\ref{fig:band-structure} (a). In the continuum low-energy model, we neglect the trigonal warping effects for simplicity, i.e. set as nontrivial a nearest-neighbor hopping within a single layer $\gamma_0 \neq 0$ and interlayer hopping between $B_{i}$ and $A_{i+1}$ sites $\gamma_1 \neq 0$. Other hopping parameters in the Slonczewski–Weiss–McClure (SWM) notations \cite{PhysRev.109.272,PhysRev.119.606,PhysRev.108.612} are set to zero: $\gamma_{i}=0$ for $ i \ge 2$. Therefore, within the \(k \cdot p\) framework, we have the following Hamiltonian density:
\begin{equation}
\label{eq:Nlayer-graphene-model}
H(\mathbf{k})  = \left( \begin{array}{ccccccc}
\varepsilon_{A_{1} } & v_F \pi^{\dagger} & 0 &  ...&0 &0 &0\\
v_F \pi& \varepsilon_{B_{1} } & \gamma_1 &  ... &0 &0 &0\\
0 & \gamma_1 &  \varepsilon_{A_{2} } & ... &0& 0 &0  \\
...&...&...&...&...&...&...\\
0& 0& 0  &...&  \varepsilon_{B_{N-1}} & \gamma_1 &0 \\
0& 0& 0  &...& \gamma_1 &  \varepsilon_{A_{N} } & v_F \pi^{\dagger} \\
0& 0 & 0 & ...  &0& v_F \pi & \varepsilon_{B_{N} } \\
\end{array} \right) ,
\end{equation}
where $\pi= \tau  k_x +i k_y$, $\tau=\pm$ are valley indices for each $\mathbf{K}_{\tau}$ valley, $v_F$ is the Fermi velocity in monolayer graphene. Its schematic lattice structure is shown in Fig.~\ref{fig:band-structure}(a).

To derive the effective two-band Hamiltonian, we employ the method described in Refs.~\onlinecite{PhysRevB.82.035409,PhysRevB.77.045429,PhysRevB.88.245445}. We consider the Green's function for Eq.~(\ref{eq:Nlayer-graphene-model}) and separate blocks corresponding to the low-energy $\psi_{\text{low}}=\left( \psi_{A_{1} }, \psi_{B_{N} } \right)^{T}$ and dimer $\psi_{\text{dim}} = \left(\psi_{B_{1}}, \psi_{A_{2}}, ...,\psi_{B_{N-1} },\psi_{A_{N} } \right)^{T}$ components:
\begin{equation}
G=\left( H-\varepsilon \, \mathbb{I}_{2N} \right)^{-1}= \left( \begin{array}{cc}
H_{11} -\varepsilon \, \mathbb{I}_{2} & H_{12}\\
H_{21} & H_{22} -\varepsilon \, \mathbb{I}_{2N-2} \\
\end{array} \right)^{-1},
\end{equation}
where $H_{11}$ is the $2\times2$ low-energy block, and $H_{22}$ is the $(2N-2) \times (2N-2)$ high-energy block, and $\mathbb{I}_{k}$ is the identity matrix of order $k$. 
The equation for the low-energy wave function is given by $\left( G^{-1}\right)_{11} \psi_{\text{low}}=0$, or explicitly
\begin{equation}
\left[ \left( H_{11} -\varepsilon \, \mathbb{I}_{2}  \right) - H_{12} \left( H_{22} -\varepsilon\, \mathbb{I}_{2N-2}  \right)^{-1} H_{21}\right] \psi_{\text{low}}=0. 
\end{equation}
The effective low-energy Hamiltonian can be obtained by expanding the equation for the low-energy wave function $\left( G^{-1}\right)_{11} \psi_{\text{low}}=0$ to the first order in $\varepsilon/\gamma_1$. Therefore, we obtain $\left( H_{\text{eff}} -\varepsilon \, \mathbb{I}_{2}  \right)  \psi_{\text{low}}=0$ with
\begin{equation}
\begin{aligned}
H_{\text{eff}}  &= \left[ \mathbb{I}_{2} + H_{12} \left(H_{22} \right)^{-2} H_{21}\right]^{-1} \\
&\times \left[ H_{11} -H_{12} \left(H_{22} \right)^{-1} H_{21}\right].
\end{aligned}
\end{equation}

In our approximation, we set the displacement field $U$ that corresponds to interlayer asymmetry between the two layers to zero. Hence, the effective two-band Hamiltonian reads as
\begin{equation}
\label{eq:eff-hamiltonian}
\begin{aligned}
H_{\text{eff}}(\mathbf{k};\tau) &=\left( -1\right)^{N-1} g_N \tau^N \left( \begin{array}{cc}
0 & \left(\pi^{\dagger}\right)^N \\
\left(\pi \right)^N & 0
\end{array} \right) \\
&=\left( -1\right)^{N-1} g_N k^{N} \left[ \cos(N\phi) \sigma_{x}+ \tau \sin(N\phi) \sigma_{y}\right],
\end{aligned} 
\end{equation}
where $g_N= \gamma_1\left( v_F / \gamma_1 \right)^{N}$~\cite{McCann_2013}. The validity of the low-energy model is discussed in Refs.~\onlinecite{PhysRevB.80.165409,PhysRevB.88.245445}. In this case, the low-energy bands of R$N$G are separated from the high-energy bands with the gap $\gamma_1$ if the trigonal warping effects are neglected. Therefore,  we consider our model valid for energies $E\lesssim \gamma_1$. The two-component spinor field $\psi_{\tau,s}$ carries the valley $\tau=\pm$ and spin $s=\pm$ indices. In the rhombohedral $N$-layer graphene, the low-energy wave functions for different valleys have the following structure: $\psi_{\tau,s}= \left( \psi_{\tau, A_{1} }, \psi_{\tau, B_{N} }  \right)^{T}_s$ for $\mathbf{K}_{\tau}$ valley. The dispersion relation reads as
\begin{equation}
\label{eq:dispresion}
\begin{aligned}
\varepsilon_{\mathbf{k}}&= \pm g_N k^{N}.  
\end{aligned}
\end{equation}
We plot the dispersion relation for four different values of $N$ in Fig.~\ref{fig:band-structure}(b). It is evident that the bands flatten at small $k$ as the number of layers increases. 

The rhombohedral $N$-layer graphene can be described by the following  low-energy Hamiltonian
\begin{equation}   
\hat{H}=\sum_{\mathbf{k},l} \sum_{i,j} \hat{c}^{\dagger}_{l,i}\left( \mathbf{k}\right)   \left[ H_{\text{eff}}(\mathbf{k};\tau) \right]^{ij}       \hat{c}_{l,j}\left( \mathbf{k}\right) +\hat{H}_{\text{int}},
\label{eq:hamiltonina-total-sq}
\end{equation}
where  $\hat{c}^{\dagger}_{l,i}\left( \mathbf{k}\right) $ ($\hat{c}_{l,i}\left( \mathbf{k}\right) $) is the creation (annihilation) operator of an electron with momentum $\mathbf{k}$ at a state $l=(\tau,\sigma)$ on a sublattice site $i=\left(A_1,B_N \right)$. We treat emergent correlations using the following density-density interaction Hamiltonian~\cite{PhysRevB.69.104504,PhysRevB.75.224509, PhysRevB.106.155115, PhysRevLett.132.186401, PhysRevB.110.045427}:
\begin{equation}  
\label{eq:int}
\hat{H}_{\text{int}}=
V_{0} \left(n_{\uparrow \mathbf{K}_{+}} n_{\downarrow \mathbf{K}_{+}} +n_{\uparrow \mathbf{K}_{-}} n_{\downarrow \mathbf{K}_{-}} \right)+V_{1} n_{\mathbf{K}_{+}}  n_{ \mathbf{K}_{-}} ,
\end{equation}
where $V_{0}>0$ and $V_{1}>0$ are parameterizing the intra- and intervalley effectively repulsive interactions, $n_{\sigma \mathbf{K}_{\tau}}$ is the number of particles with spin $\sigma$ for $\mathbf{K}_{\tau}$ valley, $n_{\mathbf{K}_{\tau}} =n_{\uparrow \mathbf{K}_{\tau}} + n_{\downarrow \mathbf{K}_{\tau}} $.
The number of particles $n_{\sigma \mathbf{K}_{\tau}}$ is defined as
\begin{equation}   
\begin{aligned} 
n_{\sigma \mathbf{K}_{\tau}}&=\sum_{\mathbf{k}}  \sum_{i}  \hat{c}^{\dagger}_{l,i}\left( \mathbf{k}\right)     \hat{c}_{l,i}\left( \mathbf{k}\right).
\label{eq:particle-number}
\end{aligned}
\end{equation}
Therefore, the interaction Hamiltonian can be expressed using the bare vertex function $V$: 
\begin{equation}
\hat{H}_{\text{int}}=\sum_{\mathbf{k},\mathbf{k}^{\prime},\mathbf{q}}   \hat{H}_{\text{int}}\left( \mathbf{k},\mathbf{k}^{\prime},\mathbf{q}\right),
\end{equation}
where
\begin{equation}
\label{eq:H_interaction}
\begin{aligned}
&\hat{H}_{\text{int}}\left( \mathbf{k},\mathbf{k}^{\prime},\mathbf{q}\right)=\frac{1}{4}\sum_{l_1l_2l_3l_4} \sum_{i,j}V_{l_1l_2l_3l_4}\left( \mathbf{k},\mathbf{k}^{\prime},\mathbf{q}\right) \\
\times &
\hat{c}^{\dagger}_{l_1,i}\left( \mathbf{k}\right)
\hat{c}^{\dagger}_{l_3,j}\left( \mathbf{k}'+\mathbf{q}\right)
\hat{c}_{l_4,j}\left( \mathbf{k}'\right)
\hat{c}_{l_2,i}\left( \mathbf{k}+\mathbf{q}\right) .
\end{aligned}
\end{equation}
After substituting Eq.~(\ref{eq:particle-number}) into Eq.~(\ref{eq:int}), we obtain
\begin{equation}
V_{l_1l_2 l_3l_4}\left( \mathbf{k},\mathbf{k}^{\prime},\mathbf{q}\right) = V_{l_1l_2,l_3l_4} \delta_{\mathbf{q},0} . \\
\end{equation}
Using the Fierz identity $\delta_{i l} \delta_{kj} = \left[ \delta_{i j} \delta_{k l} +  \left( \bm{\sigma}_{i j} \cdot \bm{\sigma}_{k l} \right) \right]/2$~\cite{10.1119/1.2074087}, function $V_{l_1l_2 l_3l_4}$ can be expressed as
\begin{equation}
\label{eq:bare_vertex}
\begin{aligned}
   V_{l_1 l_2 l_3 l_4}&=\frac{1}{2}V^{d}_{\tau_1\tau_2 \tau_3\tau_4} \delta_{\sigma_1 \sigma_2} \delta_{\sigma_3 \sigma_4}\\
&-\frac{1}{2}V^{m}_{\tau_1\tau_2 \tau_3\tau_4}
 \left( \bm{\sigma}_{\sigma_1 \sigma_2} \cdot \bm{\sigma}_{\sigma_3 \sigma_4} \right) , \\
 \end{aligned}
\end{equation}
with the interaction matrices
\begin{equation}
    V^{d}_{\tau_1\tau_2,\tau_3\tau_4}=
    \begin{cases}
    V_0&\text{if}\quad\tau_1=\tau_2=\tau_3=\tau_4 \\
    -V_1&\text{if}\quad\tau_1=\tau_4\neq\tau_3=\tau_2\\
    2V_1&\text{if}\quad\tau_1=\tau_2\neq\tau_3=\tau_4\
    \end{cases}
\end{equation}
and
\begin{equation}
    V^{m}_{\tau_1\tau_2,\tau_3\tau_4}=
    \begin{cases}
    V_0&\text{if}\quad\tau_1=\tau_2=\tau_3=\tau_4 \\
    V_1&\text{if}\quad\tau_1=\tau_4\neq\tau_3=\tau_2
    \end{cases} .
\end{equation}
This interaction Hamiltonian corresponds to the simplified two-valley Coulomb interaction model studied in Refs.~\onlinecite{PhysRevB.69.104504,PhysRevB.75.224509,PhysRevB.103.205148,PhysRevB.108.134503}. For $V_0 \neq V_1$, Eq.~(\ref{eq:int}) describes $SU(2)\times SU(2)$-symmetric interaction, since there is an $SU(2)$-symmetric interaction within each valley. For $V_0 = V_1=V$, interaction becomes $SU(4)$-symmetric. The relations between the parameters were studied in Ref.~\onlinecite{PhysRevB.110.L201113}. It was shown that the intravalley interaction parameter is greater than the intervalley for any supercell size. Nevertheless, we consider a general case of the $SU(2)\times SU(2)$-symmetric interaction.

In general, a long-range tail of the Coulomb interaction is always present in two-dimensional systems, however, we focus only on local part of the interaction, as presented in Eq.~(\ref{eq:int}). When screening from the environment (substrate, gates, remote bands) reduces the interaction beyond a few lattice periods, so that the dominant correlation effects are local, the local interaction is valid for an effective low-energy description of the system~\cite{PhysRevLett.111.036601}. Since the approach used in our work is not intended to describe the regimes of very low densities, very weak screening, or regimes close to Wigner-like phases, where nonlocal interactions are significant, we simplify our interaction Hamiltonian to consider only the local interaction.

\section{Particle-hole and particle-particle susceptibilities}
\label{sec:susceptibility}

\subsection{General properties of the two-particle correlation functions}
\label{sec:susceptibility-gen}
For the two-particle correlation functions, we introduce the generalized susceptibility as follows~\cite{RevModPhys.90.025003,PhysRevB.86.125114,TheorMethodsSCEbook}: 
\begin{equation}
\label{eq:def-gen-susceptibility}
\begin{aligned}
\chi_{l_1l_2l_3l_4}\left( x_1,x_2,x_3,x_4 \right)
&=\sum_{ij}\Big[ G_{l_1l_2l_3l_4}^{(2),\,ij}\left( x_1,x_2,x_3,x_4 \right)\\
&-G_{l_1l_4}^{(1),\,ij}\left( x_1,x_4 \right)G_{l_3l_2}^{(1),\,ji}\left( x_3,x_2 \right) \Big],
\end{aligned}
\end{equation}
where $x=\left( \tau ;\mathbf{x} \right)$ with $\tau \in \left[0,\beta \right]$ is the imaginary time, $\beta=1/T$ is the inverse temperature, and $\mathbf{x}$ is the position. Here, the one- and two-particle Green's functions are defined as
\begin{equation}
G_{l_1l_2}^{(1),\,ij}\left( x_1,x_2 \right)= \left<  \hat{T}_{\tau}  \hat{c}^{\dagger}_{l_1,i}(x_1) \hat{c}_{l_2,j}(x_2)  \right>
\end{equation}
and
\begin{equation}
\begin{aligned}
&G_{l_1l_2l_3l_4}^{(2),\,ij}\left( x_1,x_2,x_3,x_4 \right)\\
=&  \left< \hat{T}_{\tau} \hat{c}^{\dagger}_{l_1,i}(x_1) \hat{c}_{l_2,i}(x_2) \hat{c}^{\dagger}_{l_3,j}(x_3) \hat{c}_{l_4,j}(x_4)  \right>,
\end{aligned}
\end{equation}
where $\hat{T}_{\tau}$ is the time-ordering operator for imaginary time $\tau$ and $ \left< \hat{A} \right>=\text{tr} \left( e^{-\beta H}  \hat{A} \right)/ \text{tr} \left( e^{-\beta H}  \right)$. 

The one-point Green's function, omitting the sublattice indices and implying that we deal with matrices, in the momentum space reads as follows:
\begin{equation}
\begin{aligned}
G_{l_1l_2}^{(1)}\left( i\omega_{n},\mathbf{k}_1; i\omega_{n}^{\prime},\mathbf{k}_2 \right) &=\int_{0}^{\beta}d \tau_1 \int_{0}^{\beta} d \tau_2 \sum_{\mathbf{x}_1,\mathbf{x}_2} G_{l_1l_2}^{(1)}\left( x_1,x_2 \right) \\
&\times e^{i \left( \omega_{n} \tau_1 -\omega_{n}^{\prime} \tau_2 \right)} e^{i \left(\mathbf{k_1} \mathbf{x}_1-\mathbf{k_2} \mathbf{x}_2 \right)},
\end{aligned}
\end{equation}
where $\omega_{n}=(2n+1)\pi T$ is the fermionic Matsubara frequency. Since the Hamiltonian is translationally- and time-invariant, the one-particle Green's function depends only on the difference $x_1-x_2=\left( \tau_1- \tau_2 ; \mathbf{x}_1 -\mathbf{x}_2 \right)$ and the two-particle Green's function depend on three such differences. Hence, it can be shown that the Fourier transform of the Green's function $G_{l_1l_2}^{(1)}\left( x_1,x_2 \right)=G_{l_1l_2}^{(1)}\left( x_1-x_2,0 \right)$ reads as 
\begin{equation}
\begin{aligned}
G_{l_1l_2}^{(1)}\left( i\omega_{n},\mathbf{k}_1; i\omega_{n}^{\prime},\mathbf{k}_2 \right) &=\beta \mathcal{N} G_{l_1l_2}^{(1)}\left( i\omega_{n};\mathbf{k}_1 \right) \delta_{\omega_{n},\omega_{n}^{\prime}} \delta_{\mathbf{k}_1,\mathbf{k}_2},
\end{aligned}
\end{equation}
where
\begin{equation}
\begin{aligned}
G_{l_1l_2}^{(1)}\left( i\omega_{n};\mathbf{k}_1 \right) &=  \int_{0}^{\beta}d \tau_1 \sum_{\mathbf{x}_1}  e^{i  \omega_{n} \tau_1 + i \mathbf{k_1} \mathbf{x}_1}  G_{l_1l_2}^{(1)}\left( x_1,0 \right)
\end{aligned}
\end{equation}
and $\mathcal{N}$ is the number of $\mathbf{k}$-points in the summation.

\begin{figure}[t]
\hspace{0.05in}\includegraphics[width=.475\textwidth]{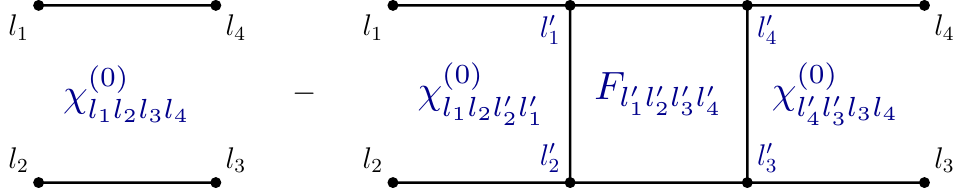}
\caption{Diagrammatic representation of Eq.~(\ref{eq:def-gen-susceptibility-FT-v1}). }
\label{fig:diagram-susceptibility}
\end{figure}

Similarly, the Fourier transform can be performed for the two-particle Green's function, where, in this case, due to the invariance of the Hamiltonian, it will depend only on three momenta and frequencies. Indeed, in the momentum space, $G^{(2)}_{l_1l_2l_3l_4}$ would be proportional to $\delta_{k_1+k_3,k_2+k_4}$, and, therefore, we can choose three pairs of parameters to define all frequencies and momenta. We denote these parameters as $k=\left( i\omega_{n}; \mathbf{k} \right)$, which corresponds to outgoing particle in the state $(\tau_{1},\sigma_{1})$, $k^{\prime}=\left( i\omega_{n}^{\prime}; \mathbf{k}^{\prime} \right)$, which corresponds to incoming particle in the state $(\tau_{4},\sigma_{4})$, and the external (bosonic) frequency and momentum $q=\left( i\Omega_{m}; \mathbf{q} \right)$, where $\Omega_{m}=2\pi m T$ is the bosonic Matsubara frequency~\cite{RevModPhys.90.025003}.
Hence, Eq.~(\ref{eq:def-gen-susceptibility}) can be rewritten in the momentum space as follows:
\begin{equation}
\label{eq:def-gen-susceptibility-FT}
\begin{aligned}
\chi_{l_1l_2l_3l_4}\left( k,k^{\prime},q \right)&=\sum_{ij} \Big[ G_{l_1l_2l_3l_4}^{(2),\,ij}\left( k,k^{\prime},q \right)\\
&-\beta \mathcal{N} G_{l_1l_4}^{(1),\,ij}\left( k \right)G_{l_3l_2}^{(1),\,ji}\left( k^{\prime}+q \right) \delta_{k,k^{\prime}} \Big] . \\
\end{aligned}
\end{equation}
The second term in Eq.~(\ref{eq:def-gen-susceptibility-FT}) corresponds to the bare susceptibility (bubble term)
\begin{equation}
\chi_{l_1l_2l_3l_4}^{(0)}\left( k,k^{\prime},q \right)=-\beta \mathcal{N} \delta_{k,k^{\prime}} \text{tr}\left[ G_{l_1l_4}^{(1)}\left( k \right)G_{l_3l_2}^{(1)}\left( k^{\prime}+q \right) \right], \\
\end{equation}
and the two-point Green's function can be expressed via the full vertex $F$, which includes all two-particle scattering events as follows:
\begin{equation}
\label{eq:def-gen-G2-FT}
\begin{aligned}
&G_{l_1l_2l_3l_4}^{(2),\,ij}\left( k,k^{\prime},q \right)=- \sum_{i_1 j_1 } G_{l_1l_1^{\prime}}^{(1),\,i i_1}\left( k \right)G_{l_2l_2^{\prime}}^{(1),\,i i_1}\left( k+q \right)\\
\times&  F_{l_1^{\prime}l_2^{\prime}l_3^{\prime}l_4^{\prime}}\left( k,k^{\prime},q \right) G_{l_3l_3^{\prime}}^{(1),\,j j_1}\left( k^{\prime}+q \right)G_{l_4l_4^{\prime}}^{(1),\,j j_1}\left( k^{\prime} \right) , \\
\end{aligned}
\end{equation}
where we have the summation over the repeating indices. Therefore, Eq.~(\ref{eq:def-gen-susceptibility-FT}) can be rewritten in terms of bare susceptibility $\chi^{(0)}$ as
\begin{equation}
\label{eq:def-gen-susceptibility-FT-v1}
\begin{aligned}
&\chi_{l_1l_2l_3l_4}\left( k,k^{\prime},q \right)=\chi_{l_1l_2l_3l_4}^{(0)}\left( k,k^{\prime},q \right)\\
-&\frac{1}{\beta^2 \mathcal{N}^2}\sum_{k^{\prime\prime},k^{\prime\prime\prime}} \chi_{l_1l_2l_2^{\prime}l_1^{\prime}}^{(0)}\left( k,k^{\prime},q \right) F_{l_1^\prime l_2^\prime l_3^\prime l_4^\prime} \left( k^{\prime\prime},k^{\prime\prime\prime},q \right) \\
\times&\chi_{l_4^\prime l_3^\prime  l_3l_4}^{(0)}\left( k^{\prime\prime\prime},k^{\prime},q \right). \\
\end{aligned}
\end{equation}
Its diagrammatic representation is shown in Fig.~\ref{fig:diagram-susceptibility}.
By introducing matrix notations and assuming that for every matrix multiplication we additionally have summation over momentum and Matsubara frequency, Eq.~(\ref{eq:def-gen-susceptibility-FT-v1}) reads as follows:
\begin{equation}
\label{eq:def-gen-susceptibility-FT-v2}
\chi=\chi^{(0)}- \chi^{(0)} F \chi^{(0)} .
\end{equation}

\begin{figure*}[t]
\includegraphics[width=.99\textwidth]{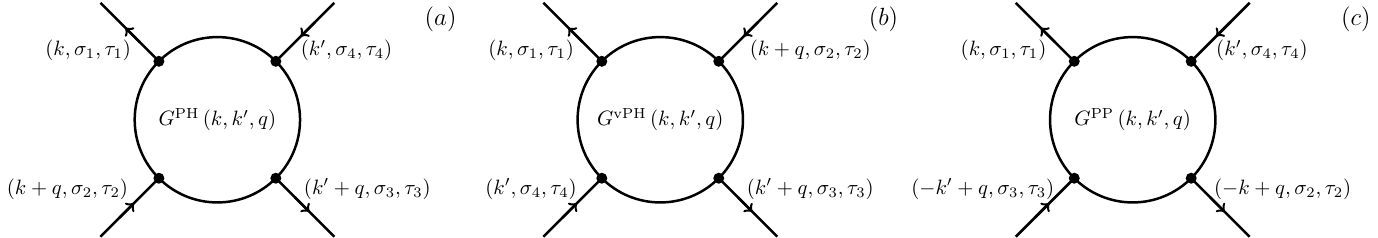}
\caption{(a) Particle-hole, (b) transverse particle-hole and (c) particle-particle channels of the two-point correlation function $G_{l_1 l_2 l_3 l_4}$.}
\label{fig:diagram-G}
\end{figure*}

Since we are considering two-particle level, the reducibility of the vertex function is related to a specific channel~\cite{PhysRevB.86.125114}. Therefore, for these Fourier-transformed functions, we introduce the particle-hole (PH), transverse (vertical) particle-hole (vPH), and particle-particle (PP) channels for the arbitrary two-particle correlation function $G_{l_1 l_2 l_3 l_4}$ as presented in Fig.~\ref{fig:diagram-G}. 

The full vertex $F$ in the Bethe-Salpeter equation for the generalized susceptibility can be represented using the parquet equation in terms of the sum of all fully irreducible vertex $\Lambda$ and the channel-reducible vertices $\Phi$ as follows~\cite{RevModPhys.90.025003,PhysRevB.86.125114}:
\begin{equation}
\label{eq:def-full-vertex}
\begin{aligned}
F&=\Lambda+\Phi^{\text{PH}}+\Phi^{\text{vPH}}+\Phi^{\text{PP}}.\\
\end{aligned}
\end{equation}
For each channel, the full vertex $F$ can be represented in terms of a channel-reducible vertex $\Phi^{r}$ and a channel-irreducible vertex $\Gamma^{r}$ in a specific channel $r=\text{PH},\,\,\text{vPH},\,\,\text{PP}$:
\begin{equation}
\label{eq:def-full-vertex-channel}
\begin{aligned}
F&=\Phi^{r}+\Gamma^{r}.\\
\end{aligned}
\end{equation}

Combining Eqs.~(\ref{eq:def-full-vertex}) and (\ref{eq:def-full-vertex-channel}), we obtain the relation between the channel-irreducible vertices $\Gamma$ and the channel-reducible vertices $\Phi$ via the following parquet equations
\begin{equation}
\label{eq:irreducible-Bethe-Salpeter}
\begin{aligned}
\Gamma^{\text{PH}}&=\Lambda^{\text{PH}}+\Phi^{\text{vPH}}+\Phi^{\text{PP}},\\
\Gamma^{\text{vPH}}&=\Lambda^{\text{vPH}}+\Phi^{\text{PH}}+\Phi^{\text{PP}},\\
\Gamma^{\text{PP}}&=\Lambda^{\text{PP}}+\Phi^{\text{PH}}+\Phi^{\text{vPH}}.\\
\end{aligned}
\end{equation}
The channel-reducible vertices are connected to bare susceptibilities and full vertices by the following relations:
\begin{equation}
\label{eq:reducible-Bethe-Salpeter}
\begin{aligned}
\Phi^{\text{PH}}&=-\Gamma^{\text{PH}}\chi^{(0),\text{PH}}F^{\text{PH}},\\
\Phi^{\text{vPH}}&=-\Gamma^{\text{vPH}}\chi^{(0),\text{vPH}}F^{\text{vPH}},\\
\Phi^{\text{PP}}&=-\frac{1}{2} \Gamma^{\text{PP}}\chi^{(0),\text{PP}}F^{\text{PP}}.
\end{aligned}
\end{equation}
where the factor $1/2$ in the PP channel must be introduced to avoid double-counting of the diagrams.
By combining Eqs.~(\ref{eq:def-full-vertex-channel}) and (\ref{eq:reducible-Bethe-Salpeter}), the Bethe-Salpeter equations for full vertices in the PH, vPH, and PP channels can be expressed as follows:
\begin{equation}
\label{eq:app_full_vertex}
\begin{aligned}
F^{\text{PH}}&=\Gamma^{\text{PH}}-\Gamma^{\text{PH}}\chi^{(0),\text{PH}}F^{\text{PH}},\\
F^{\text{vPH}}&=\Gamma^{\text{vPH}}-\Gamma^{\text{vPH}}\chi^{(0),\text{vPH}}F^{\text{vPH}},\\
F^{\text{PP}}&=\Gamma^{\text{PP}}-\frac{1}{2} \Gamma^{\text{PP}}\chi^{(0),\text{PP}}F^{\text{PP}}.
\end{aligned}
\end{equation}
From Eqs.~(\ref{eq:irreducible-Bethe-Salpeter}) and (\ref{eq:reducible-Bethe-Salpeter}), we obtain the Bethe-Salpeter equation for the channel-irreducible vertices:
\begin{equation}
\label{eq:app_ir_vertex}
\begin{aligned}
\Gamma^{\text{PH}}&=\Lambda^{\text{PH}}-\Gamma^{\text{vPH}}\chi^{(0),\text{vPH}}F^{\text{vPH}}-\frac{1}{2} \Gamma^{\text{PP}}\chi^{(0),\text{PP}}F^{\text{PP}},\\
\Gamma^{\text{vPH}}&=\Lambda^{\text{vPH}}-\Gamma^{\text{PH}}\chi^{(0),\text{PH}}F^{\text{PH}}-\frac{1}{2} \Gamma^{\text{PP}}\chi^{(0),\text{PP}}F^{\text{PP}},\\
\Gamma^{\text{PP}}&=\Lambda^{\text{PP}}-\Gamma^{\text{PH}}\chi^{(0),\text{PH}}F^{\text{PH}}-\Gamma^{\text{vPH}}\chi^{(0),\text{vPH}}F^{\text{vPH}}.
\end{aligned}
\end{equation}
Combining the Eqs.~(\ref{eq:def-gen-susceptibility-FT-v2}) and (\ref{eq:app_full_vertex}), we obtain the relation between generalized susceptibility $\chi$ and channel-irreducible vertices $\Gamma$ for each channel:
\begin{equation}
\label{eq:app_gen_susceptibility}
\begin{aligned}
\chi^{\text{PH}}&=\chi^{(0),\text{PH}}-\chi^{(0),\text{PH}}\Gamma^{\text{PH}}\chi^{\text{PH}},\\
\chi^{\text{vPH}}&=\chi^{(0),\text{vPH}}-\chi^{(0),\text{vPH}}\Gamma^{\text{vPH}}\chi^{\text{vPH}},\\
\chi^{\text{PP}}&=\chi^{(0),\text{PP}}-\frac{1}{2} \chi^{(0),\text{PP}} \Gamma^{\text{PP}} \left[ \chi^{\text{PP}}+ \chi^{(0),\text{PP}}\right].
\end{aligned}
\end{equation}
The correlated phases emerge when the generalized susceptibility has a divergence. We analyze the generalized susceptibilities using the random phase (RPA) and the parquet approximations for the vertex functions in Sec.~\ref{sec:rpa} and \ref{sec:parquet} respectively.

In the RPA, the channel irreducible vertices in Eq.~(\ref{eq:app_gen_susceptibility}) are approximated by the bare vertices $V$, given by Eq.~(\ref{eq:bare_vertex}), for calculating the generalized susceptibilities. 

In the parquet approximation, we approximate the fully irreducible vertices $\Lambda$ with bare vertices. Using Eq.~(\ref{eq:app_full_vertex}), we express the full vertices $F$ via the bare susceptibilities and the channel irreducible vertices $\Gamma$. The latter are then approximated by the fully irreducible vertices $\Lambda$. Hence, the approximated fully reducible vertices $\Phi$ in the corresponding channel in Eq.~(\ref{eq:reducible-Bethe-Salpeter}) are given by
\begin{equation}
\label{eq:app_reducible_vertex-approx}
\begin{aligned}
\Phi^{\text{PH}}&=-\Lambda^{\text{PH}} \chi^{(0)} \left[1+\Lambda^{\text{PH}} \chi^{(0),\text{PH}} \right]^{-1} \Lambda^{\text{PH}},\\
\Phi^{\text{vPH}}&=-\Lambda^{\text{vPH}} \chi^{(0)} \left[1+\Lambda^{\text{vPH}} \chi^{(0),\text{vPH}} \right]^{-1} \Lambda^{\text{vPH}},\\
\Phi^{\text{PP}}&=-\Lambda^{\text{PP}} \chi^{(0)} \left[1+\frac{1}{2} \Lambda^{\text{PP}} \chi^{(0),\text{PP}} \right]^{-1} \Lambda^{\text{PP}}.
\end{aligned}
\end{equation}
The channel irreducible vertices $\Gamma$ are obtained by substituting the approximated fully reducible vertices in the rhs of Eq.~(\ref{eq:irreducible-Bethe-Salpeter}). Then, these vertices are used in Eq.~(\ref{eq:app_gen_susceptibility}).

\subsection{Bare susceptibilities}

\label{sec:susceptibility-bare}

From the quantities given by Eq.~(\ref{eq:app_gen_susceptibility}), the physical susceptibilities can be constructed by summing over the two momenta $\mathbf{k}$ and $\mathbf{k}^{\prime}$ and two fermionic Matsubara frequencies $\omega_{n}$ and $\omega_{n}^{\prime}$. 
We set the external frequency to zero to focus on instabilities, which have infinite lifetimes according to the Kramers-Kronig relations. At the same time for simplicity, we also neglect the external momentum dependence within a single valley~\cite{PhysRevB.108.134503}.
Therefore, the bare particle-hole and particle-particle susceptibilities at zero external frequency and momentum, which can be presented as the loop diagrams, read as follows:
\begin{equation}
\begin{aligned}
\chi^{(0),\text{PH}}_{l_1 l_2 l_3 l_4}&= \delta_{\tau_1 \tau_4} \delta_{\tau_2 \tau_3}  \delta_{\sigma_1 \sigma_4} \delta_{\sigma_2 \sigma_3} \, \chi^{(0),\text{PH}}_{\tau_1,\tau_2},\\
\chi^{(0),\text{PP}}_{l_1 l_2 l_3 l_4}&= \delta_{\tau_1 \tau_4} \delta_{\tau_2 \tau_3}  \delta_{\sigma_1 \sigma_4} \delta_{\sigma_2 \sigma_3} \, \chi^{(0),\text{PP}}_{\tau_1,\tau_2},\\
\end{aligned}
\end{equation}
where
\begin{equation}
\label{eq:PH-0-def}
\chi^{(0),\text{PH}}_{\tau_1,\tau_2}=-\frac{1}{\beta \mathcal{N}} \sum_{\omega_n}\sum_{\mathbf{k}} \text{tr} \left[ G_{\tau_1}(i \omega_n;\mathbf{k})G_{\tau_2}(i \omega_n;\mathbf{k}) \right],
\end{equation}
\begin{equation}
\label{eq:PP-0-def}
\chi^{(0),\text{PP}}_{\tau_1,\tau_2}=-\frac{1}{\beta \mathcal{N}} \sum_{\omega_n}\sum_{\mathbf{k}}\text{tr} \left[ G_{\tau_1}(i \omega_n;\mathbf{k})G_{\tau_2}(-i \omega_n;-\mathbf{k}) \right].
\end{equation}

The susceptibilities can be rewritten in terms of the spectral function $A_{\tau}\left( \omega;\mathbf{k} \right)$; see Appendix~\ref{app:Kubo} for details. For generality, the expressions in the form of numerical quadrature for arbitrary values of $\mu$, $T$, and $U$ are given by Eqs.~(\ref{eq:intra-ph-susceptibility})-(\ref{eq:inter-pp-susceptibility}).

For a zero displacement field $U$, we can obtain simplified analytical expressions. The bare particle-hole and particle-particle susceptibilities are given by
\begin{equation}
\label{eq:susceptibilities-via-tilde}
\begin{aligned}
\chi^{(0),\text{PH}}_{\tau,\tau} &=\tilde{\chi} , \\
\chi^{(0),\text{PH}}_{\tau,-\tau}  &= \frac{N-1}{N-2} \tilde{\chi} , \\
\chi^{(0),\text{PP}}_{\tau,\tau}  &=-\frac{N-1+(-1)^N}{N-2} \tilde{\chi} ,\\
\chi^{(0),\text{PP}}_{\tau,-\tau}  &=- \frac{N-1}{N-2}  \tilde{\chi} ,
\end{aligned}
\end{equation}
where we introduced the function $\tilde{\chi}=\tilde{\chi}(\beta,\mu)$ as follows:
\begin{equation}
\label{eq:tilde-chi-def}
\begin{aligned}
\tilde{\chi} &=-\frac{\beta }{2\pi N \left( \beta g_N \right)^{\frac{2}{N}}}   \Gamma \left(\frac{2}{N}\right) \sum_{\lambda = \pm} \text{Li}_{\frac{2}{N}-1}\left(-e^{\lambda \beta \mu }\right) , \\
\end{aligned}
\end{equation}
with $\text{Li}_{n}(x)$ being the polylogarithm function.

Later, we will use the matrix form for bare susceptibility in the following representation in the PH channel
\begin{equation}
\begin{aligned}
\chi^{(0),\text{PH}}&= \text{diag} \left( \chi_{+,+}^{(0),\text{PH}},\, \chi_{+,-}^{(0),\text{PH}},\, \chi_{-,+}^{(0),\text{PH}},\, \chi_{-,-}^{(0),\text{PH}} \right), \\
\end{aligned}
\end{equation}
and in the PP channel
\begin{equation}
\begin{aligned}
\chi^{(0),\text{PP}}&= \text{diag} \left( \chi_{+,+}^{(0),\text{PP}},\, \chi_{+,-}^{(0),\text{PP}},\, \chi_{-,+}^{(0),\text{PP}},\, \chi_{-,-}^{(0),\text{PP}} \right). \\
\end{aligned}
\end{equation}
Such ordering corresponds to the structure of the channel-irreducible vertices used in Ref.~\onlinecite{PhysRevB.69.104504}.

\section{Random Phase Approximation}
\label{sec:rpa}

Since interaction Hamiltonian given by Eq.~(\ref{eq:int}) has $SU(2)\times SU(2)$ symmetry, it is not necessary to calculate the two-point correlation function for every spin combination $\left\{ \sigma_{i} \right\}$ because other combinations may be reproduced using symmetry relations or may equal to zero due to spin conservation. Hence, as discussed in Ref.~\onlinecite{PhysRevB.86.125114}, an arbitrary two-particle correlation function $G_{\sigma_1\sigma_2\sigma_3\sigma_4}$ can be described by four independent channels: magnetic ($m$), density ($d$), singlet ($s$), and triplet ($t$), which are defined as follows:
\begin{equation}
\label{eq:spin-channel-sep}
\begin{aligned}
G^{m}&=G^{\text{PH}}_{\uparrow\uparrow\downarrow\downarrow}-G^{\text{PH}}_{\uparrow\uparrow\uparrow\uparrow},\\
G^{d}&=G^{\text{PH}}_{\uparrow\uparrow\uparrow\uparrow}+G^{\text{PH}}_{\uparrow\uparrow\downarrow\downarrow},\\
G^{s}&=G^{\text{PP}}_{\uparrow\uparrow\downarrow\downarrow}-G^{\text{PP}}_{\uparrow\downarrow\downarrow\uparrow},\\
G^{t}&=G^{\text{PP}}_{\uparrow\uparrow\downarrow\downarrow}+G^{\text{PP}}_{\uparrow\downarrow\downarrow\uparrow}.\\
\end{aligned}
\end{equation}

Therefore, we separate the generalized susceptibilities and vertices in these channels by spin symmetry. The quantities $\chi^{m,d,s,t}$, $\Lambda^{m,d,s,t}$, $\Gamma^{m,d,s,t}$, $F^{m,d,s,t}$ are defined in accordance with Eq.~(\ref{eq:spin-channel-sep}). 

In these channels, we define the ordered states with the corresponding phase-ordering operators 
\begin{equation}
\hat{\Phi}=\sum_{\mathbf{k},\mathbf{k}^{\prime}} \sum_{i}  \hat{c}_{l,i}^{\dagger}(\mathbf{k})\left[M_{\Phi}\left(\mathbf{k},\mathbf{k}^{\prime} \right)\right]^{l l^{\prime}} \hat{c}_{l^{\prime},i} (\mathbf{k}^{\prime}) 
\end{equation}
as presented in Table.~\ref{tab:table-orders}.
By comparing the eigenvectors of the matrices $\chi^{(0),\text{PH}}\Gamma^{m}$, $\chi^{(0),\text{PH}}\Gamma^{d}$, $ \chi^{(0),\text{PP}}\Gamma^{s}/2$, $\chi^{(0),\text{PP}}\Gamma^{t}/2$ with the eigenvectors of the phase-ordering operators $M_{\Phi}\left(\mathbf{k},\mathbf{k}^{\prime} \right)$, we identify the corresponding correlated phase. Therefore, the order parameters corresponding to the IVC state in the density and magnetic channel are the same as in charge density wave (CDW) and spin density wave (SDW) states, respectively. Both CDW and SDW break translational symmetry of the system. In contrast, the Stoner phase corresponds to a spatially uniform symmetry-breaking phase.

\begin{table*}[t]
\caption{\label{tab:table-orders} Summary of correlated phases and the corresponding phase-ordering operators $M_{\Phi}\left(\mathbf{k},\mathbf{k}^{\prime} \right)$. Here, $s_{i}$ and $\tau_{i}$ are the Pauli matrices in the spin and valley spaces, respectively.}
\begin{ruledtabular}
\begin{tabular}{lc}
Correlated phase &  Operator $M_{\Phi}$\\ \hline
Stoner instability in the magnetic channel
or intravalley ferromagnetism &  $
M_{\Phi}^{\text{St}\,(m)}  = s_{x,y,z} \otimes \left( \tau_{0} \pm \tau_{z} \right) \delta (\mathbf{k} - \mathbf{k}^{\prime})$
 \\
 IVC phase in the magnetic channel &  $
M_{\Phi}^{\text{IVC}\,(m)}  = s_{x,y,z} \otimes \left( \tau_{x} \pm i \tau_{y} \right) \delta (\mathbf{k} - \mathbf{k}^{\prime})$
 \\
 Stoner instability in the density channel or valley polarization &  $
M_{\Phi}^{\text{St}\,(d)}  = s_{0} \otimes  \tau_{z} \delta (\mathbf{k} - \mathbf{k}^{\prime})$
 \\
IVC phase in the density channel &  $
M_{\Phi}^{\text{IVC}\,(d)} = s_{0} \otimes   \left( \tau_{x} \pm i \tau_{y} \right)  \delta (\mathbf{k} - \mathbf{k}^{\prime} ) $
 \\
 Cooper instability in the singlet channel &  $M_{\Phi}^{\text{Cooper}\,(s)}  = s_{x,y,z}  \otimes  \tau_{x} \,  \delta (\mathbf{k} + \mathbf{k}^{\prime}) $
 \\
Cooper instability in the triplet channel&  $M_{\Phi}^{\text{Cooper}\,(t)} =   s_{0}  \otimes \tau_{y} \, \delta (\mathbf{k} + \mathbf{k}^{\prime}) $ \\
Pair-density wave in the singlet channel &  $M_{\Phi}^{\text{PDW}} = s_{x,y,z}  \otimes \left(\tau_{0}\pm\tau_{z}\right) \delta (\mathbf{k} + \mathbf{k}^{\prime} )$
 \\
\end{tabular}
\end{ruledtabular}
\end{table*}

To quantitatively describe the correlated phases in the random phase approximation, we approximate the channel-irreducible vertices $\Gamma$ with the bare vertex $V$ in a corresponding channel in Eq.~(\ref{eq:app_gen_susceptibility}) as stated previously in Sec.~\ref{sec:susceptibility-gen}.

\subsection{Magnetic channel}
In the magnetic channel, the generalized susceptibility $\chi^{m}=\chi^{\text{PH}}_{\uparrow\uparrow\downarrow\downarrow}-\chi^{\text{PH}}_{\uparrow\uparrow\uparrow\uparrow}$ obeys the following equation:
\begin{equation}
\label{eq:m-gen-susceptibility}
\begin{aligned}
\chi^{m}&= \chi^{(0),\text{PH}}_{\uparrow\uparrow\downarrow\downarrow}-\chi^{(0),\text{PH}}_{\uparrow\uparrow\uparrow\uparrow}\\
&-\sum_{\left\{ \sigma_i \right\}}\chi^{(0),\text{PH}}_{\uparrow\uparrow\sigma_1\sigma_2} \Gamma^{\text{PH}}_{\sigma_2 \sigma_1 \sigma_4 \sigma_3}\chi^{\text{PH}}_{\sigma_3 \sigma_4 \downarrow\downarrow}\\
&+\sum_{\left\{ \sigma_i \right\}}\chi^{(0),\text{PH}}_{\uparrow\uparrow\sigma_1\sigma_2} \Gamma^{\text{PH}}_{\sigma_2 \sigma_1 \sigma_4 \sigma_3}\chi^{\text{PH}}_{\sigma_3 \sigma_4 \uparrow\uparrow} \\
&=-\chi^{(0),\text{PH}}+\chi^{(0),\text{PH}} \Gamma^m \chi^{m},
\end{aligned}
\end{equation}
where we used the notation $\Gamma^{m}=\Gamma^{\text{PH}}_{\uparrow\uparrow\downarrow\downarrow}-\Gamma^{\text{PH}}_{\uparrow\uparrow\uparrow\uparrow}$. 

The full vertex $F^{m}=F^{\text{PH}}_{\uparrow\uparrow\downarrow\downarrow}-F^{\text{PH}}_{\uparrow\uparrow\uparrow\uparrow}$ reads as follows:
\begin{equation}
\label{eq:m-F-vertex}
\begin{aligned}
F^{m}&=\Gamma^{\text{PH}}_{\uparrow\uparrow\downarrow\downarrow} - \Gamma^{\text{PH}}_{\uparrow\uparrow\uparrow\uparrow} \\
&-  \sum_{\left\{ \sigma_i \right\}} \Gamma^{\text{PH}}_{\uparrow\uparrow\sigma_1\sigma_2} \chi^{(0),\text{PH}}_{\sigma_2 \sigma_1 \sigma_4 \sigma_3} F^{\text{PH}}_{\sigma_3 \sigma_4 \downarrow\downarrow}\\
&+ \sum_{\left\{ \sigma_i \right\}} \Gamma^{\text{PH}}_{\uparrow\uparrow\sigma_1\sigma_2} \chi^{(0),\text{PH}}_{\sigma_2 \sigma_1 \sigma_4 \sigma_3} F^{\text{PH}}_{\sigma_3 \sigma_4 \uparrow\uparrow}  \\
&=\Gamma^{m} + \Gamma^{m}\chi^{(0),\text{PH}} F^{m}.
\end{aligned}
\end{equation}
Therefore, in the magnetic channel, we obtain the following expression for generalized susceptibility by combining Eqs.~(\ref{eq:m-gen-susceptibility}), (\ref{eq:m-F-vertex}), and (\ref{eq:def-gen-susceptibility-FT-v2}):
\begin{equation}
\label{eq:gen-susc-m-def}
\chi^{m}=-\left[ \mathbb{I}-\chi^{(0),\text{PH}}\Gamma^{m} \right]^{-1}\chi^{(0),\text{PH}},
\end{equation}
where the approximated vertex $\Gamma^{m}$ reads as
\begin{equation}
\Gamma^{m}\simeq
\left(\begin{array}{cccc}
V_0&0&0&0\\
0&V_1&0&0\\
0&0&V_1&0\\
0&0&0&V_0
\end{array}\right),
\end{equation}
with $\mathbb{I}$ being the $4 \times 4$ identity matrix.

From Eq.~(\ref{eq:gen-susc-m-def}), we can see that the correlated phases emerge when the eigenvalues of the equation $ \chi^{(0),\text{PH}}\Gamma^{m} \phi =\lambda \phi$ equal to one. Hence, for intravalley transitions that correspond to the Stoner instability, we have the following equation for the phase transition:
\begin{equation}
\label{M-phase-tr-Stoner}
\lambda^{\text{Stoner}\,(m)}=V_0 \chi_{\tau,\tau}^{(0),\text{PH}} =1.
\end{equation}
The eigenvalue $\lambda^{\text{Stoner}\,(m)}$ corresponds to the following eigenvectors: $\phi=\left(1,0,0,0\right)^{T}$ and $\phi=\left(0,0,0,1\right)^{T}$.

For intervalley transitions, the intervalley coherent (IVC) phase emerges when its correspondent eigenvalue equals to one:
\begin{equation}
\label{M-phase-tr-IVC}
\lambda^{\text{IVC}\,(m)}=V_1 \chi_{\tau,-\tau}^{(0),\text{PH}}=1.
\end{equation}
In this case, the eigenvectors are $\phi=\left( 0,1,0,0 \right)^{T}$ and $\phi=\left( 0,0,1,0 \right)^{T}$.

Equation~(\ref{eq:PH-0-def}) shows that the total momentum transfer $\mathbf{Q}$, including the momentum of the $\mathbf{K}_{\tau}$ point, for the intravalley susceptibility $\chi_{\tau,\tau}^{(0),\text{PH}}$ is zero since $\mathbf{Q}=\left( \mathbf{k} + \mathbf{K}_{\tau}\right)-\left( \mathbf{k} + \mathbf{K}_{\tau}\right)=\mathbf{0}$. Therefore, it is expected that $\chi_{\tau,\tau}^{\text{PH}}$ corresponds to the Stoner instability. The total momentum transfer $\mathbf{Q}$ of the intervalley susceptibility $\chi_{\tau,-\tau}^{(0),\text{PH}}$ is nonzero since $\mathbf{Q}=\left( \mathbf{k} + \mathbf{K}_{\tau}\right)-\left( \mathbf{k} + \mathbf{K}_{-\tau}\right)=2\mathbf{K}_{\tau}$. Hence, it corresponds to the intervalley coherent phase.

Eqs.~(\ref{M-phase-tr-Stoner}) and (\ref{M-phase-tr-IVC}) define  the critical temperature $T_c=1/\beta_{c}$ as a function of $\mu$ and other model parameters. In particular, at the charge neutrality point ($\mu=0$), the critical temperature $T_c$, at which the phase transition occurs, can be determined analytically. We have the correlated phase for $T < T_c$. In the case of $SU(2)\times SU(2)$-symmetric interaction, the critical temperature for the Stoner instability in the magnetic channel is given by
\begin{equation}
\label{eq:rpa-t-st-m}
\begin{aligned}
T_c^{\text{Stoner}\,(m)}&=V_0^{\frac{N}{N-2}} \gamma_1^{\frac{2-2N}{2-N}} v_F^{\frac{2N}{2-N}}\\
&\times\left[\Gamma \left(\frac{2}{N}\right) \frac{1-2^{2-\frac{2}{N}}}{\pi N} \zeta\left(\frac{2}{N}-1\right) \right]^{\frac{N}{N-2} } , \\
\end{aligned}
\end{equation}
where $\zeta(x)$ is a Riemann zeta function. 
Similarly, for the IVC phase in the magnetic channel, we obtain
\begin{equation}
\label{eq:rpa-t-ivc-m}
\begin{aligned}
 T_c^{\text{IVC}\,(m)}&=\left( \frac{ N-1}{N-2} V_1 \right)^{\frac{N}{N-2}} \gamma_1^{\frac{2-2N}{2-N}} v_F^{\frac{2N}{2-N}}\\
 &\times\left[  \Gamma \left(\frac{2}{N}\right) \frac{1-2^{2-\frac{2}{N}}}{\pi N} \zeta\left(\frac{2}{N}-1\right) \right]^{\frac{N}{N-2} } .
\end{aligned}
\end{equation}
The critical temperatures in Eqs.~(\ref{eq:rpa-t-st-m}) and (\ref{eq:rpa-t-ivc-m}) monotonically increase with the increase of the number of layers.

\subsection{Density channel}

For generalized susceptibility in the density channel $\chi^{d}=\chi^{\text{PH}}_{\uparrow\uparrow\uparrow\uparrow}+\chi^{\text{PH}}_{\uparrow\uparrow\downarrow\downarrow}$, we obtain the following equation:
\begin{equation}
\label{eq:d-gen-susceptibility}
\begin{aligned}
\chi^{d}&=\chi^{(0),\text{PH}}_{\uparrow\uparrow\uparrow\uparrow}+ \chi^{(0),\text{PH}}_{\uparrow\uparrow\downarrow\downarrow}\\
&-\sum_{\left\{ \sigma_i \right\}}\chi^{(0),\text{PH}}_{\uparrow\uparrow\sigma_1\sigma_2} \Gamma^{\text{PH}}_{ \sigma_2\sigma_1 \sigma_4 \sigma_3}\chi^{\text{PH}}_{\sigma_3 \sigma_4 \uparrow\uparrow} \\
&   -\sum_{\left\{ \sigma_i \right\}}\chi^{(0),\text{PH}}_{\uparrow\uparrow\sigma_1\sigma_2} \Gamma^{\text{PH}}_{ \sigma_2 \sigma_1\sigma_4 \sigma_3}\chi^{\text{PH}}_{\sigma_3 \sigma_4 \downarrow\downarrow}\\
&=\chi^{(0),\text{PH}}-\chi^{(0),\text{PH}} \Gamma^d \chi^{d}
\end{aligned}
\end{equation}
with $\Gamma^{d}=\Gamma^{\text{PH}}_{\uparrow\uparrow\uparrow\uparrow}+\Gamma^{\text{PH}}_{\uparrow\uparrow\downarrow\downarrow}$.
For the full vertex $F^{d}=F^{\text{PH}}_{\uparrow\uparrow\uparrow\uparrow}+F^{\text{PH}}_{\uparrow\uparrow\downarrow\downarrow}$, we arrive at
\begin{equation}
\label{eq:d-F-vertex}
\begin{aligned}
F^{d}&=\Gamma^{\text{PH}}_{\uparrow\uparrow\uparrow\uparrow}+\Gamma^{\text{PH}}_{\uparrow\uparrow\downarrow\downarrow} \\
&- \sum_{\left\{ \sigma_i \right\}} \Gamma^{\text{PH}}_{\uparrow\uparrow\sigma_1\sigma_2} \chi^{(0),\text{PH}}_{\sigma_2 \sigma_1 \sigma_4 \sigma_3} F^{\text{PH}}_{\sigma_3 \sigma_4 \uparrow\uparrow} \\
&- \sum_{\left\{ \sigma_i \right\}} \Gamma^{\text{PH}}_{\uparrow\uparrow\sigma_1\sigma_2} \chi^{(0),\text{PH}}_{\sigma_2 \sigma_1 \sigma_4 \sigma_3} F^{\text{PH}}_{\sigma_3 \sigma_4 \downarrow\downarrow} \\
&=\Gamma^{d}- \Gamma^{d}\chi^{(0),\text{PH}} F^{d}.
\end{aligned}
\end{equation}

Similarly to Eq.~(\ref{eq:gen-susc-m-def}), in the density channel, the generalized susceptibility reads as follows:
\begin{equation}
\label{eq:gen-susc-d-def}
\chi^{d}=\left[ \mathbb{I}+\chi^{(0),\text{PH}}\Gamma^{d} \right]^{-1}\chi^{(0),\text{PH}},\\
\end{equation}
with approximated vertex $\Gamma^{d}$ as
\begin{equation}
\Gamma^{d}\simeq
\left(\begin{array}{cccc}
V_0&0&0&2V_1\\
0&-V_1&0&0\\
0&0&-V_1&0\\
2V_1&0&0&V_0
\end{array}\right).
\end{equation}

From Eq.~(\ref{eq:gen-susc-d-def}), we obtain the equation for the phase transitions, when the eigenvalues of the equation $-\chi^{(0),\text{PH}}\Gamma^{d} \phi =\lambda \phi$ equal to one. Hence, for the Stoner instability in the density channel, we have the following equation:
\begin{equation}
\label{D-phase-tr-Stoner}
\lambda^{\text{Stoner}\,(d)} = \left(2V_1-V_0\right) \chi_{\tau,\tau}^{(0),\text{PH}}=1 .
\end{equation}
In this case, the eigenvalue  $\lambda^{\text{Stoner},\,(d)}$ corresponds to the eigenvector $\phi=\left( 1,0,0,- 1\right)^{T}$. 

The equation for the IVC phase transition in the density channel coincides with Eq.~(\ref{M-phase-tr-IVC}) in the magnetic channel:
\begin{equation}
\label{D-phase-tr-IVC}
\lambda^{\text{IVC}\,(d)}=V_1 \chi_{\tau,-\tau}^{(0),\text{PH}}=1 .
\end{equation}
The eigenvectors are also the same as in the magnetic channel: $\phi=\left( 0,1,0,0 \right)^{T}$ and $\phi=\left( 0,0,1,0 \right)^{T}$.

At the charge neutrality point, in the case of $SU(2)\times SU(2)$-symmetric interaction, we obtain the following critical temperature for the Stoner instability in the density channel:
\begin{equation}
\label{eq:rpa-t-st-d}
\begin{aligned}
T_c^{\text{Stoner}\,(d)}&=\left(2V_1-V_0 \right)^{\frac{N}{N-2} } \gamma_1^{\frac{2-2N}{2-N}} v_F^{\frac{2N}{2-N}}  \\
&\times \left[ \Gamma \left(\frac{2}{N}\right) \frac{1-2^{2-\frac{2}{N}}}{\pi N} \zeta\left(\frac{2}{N}-1\right) \right]^{\frac{N}{N-2} }.
\end{aligned}
\end{equation}
Furthermore, Eq.~(\ref{D-phase-tr-Stoner}) shows that a solution for $T_{c}^{\text{Stoner}\,(d)}$ exists only when $V_{1} > V_{0}/2$ since $\tilde{\chi}$ is positive for any $\mu$. Otherwise, there is no solution for the Stoner instability in the density channel. The critical temperature $T^{\text{IVC}\,(d)}$ for the IVC phase in the density channel coincides with $T^{\text{IVC}\,(m)}$ in Eq.~(\ref{eq:rpa-t-ivc-m}). Hence, in what follows, we will drop indices $(m)$ and $(d)$ for the IVC phase.   

\subsection{Singlet and triplet channels}

Similarly, we perform the separation of the generalized susceptibilities and vertices by spin symmetry in the singlet and triplet channels. For the susceptibility in the singlet channel $\chi^{s}=\chi^{\text{PP}}_{\uparrow\uparrow\downarrow\downarrow}-\chi^{\text{PP}}_{\uparrow\downarrow\downarrow\uparrow} $, we obtain 
\begin{equation}
\label{eq:s-gen-susceptibility}
\begin{aligned}
\chi^{s}&=\chi^{(0),\text{PP}}_{\uparrow\uparrow\downarrow\downarrow} - \chi^{(0),\text{PP}}_{\uparrow\downarrow\downarrow\uparrow} \\
&-\frac{1}{2} \sum_{ \left\{ \sigma_i \right\} }\chi^{(0),\text{PP}}_{\uparrow\sigma_1\downarrow\sigma_2} \Gamma^{\text{PP}}_{\sigma_2 \sigma_4  \sigma_1 \sigma_3}\left( \chi^{(0),\text{PP}}_{\sigma_3\uparrow  \sigma_4\downarrow } + \chi^{\text{PP}}_{\sigma_3\uparrow  \sigma_4 \downarrow}  \right) \\
&+\frac{1}{2} \sum_{\left\{ \sigma_i \right\}}\chi^{(0),\text{PP}}_{\uparrow\sigma_1\downarrow \sigma_2} \Gamma^{\text{PP}}_{\sigma_2 \sigma_4 \sigma_1 \sigma_3 }\left( \chi^{(0),\text{PP}}_{\sigma_3\downarrow \sigma_4 \uparrow   } + \chi^{\text{PP}}_{\sigma_3\downarrow \sigma_4 \uparrow   }  \right)\\ 
&=-\chi^{(0),\text{PP}}-\frac{1}{2}  \chi^{(0),\text{PP}} \Gamma^{s} \left(  \chi^{(0),\text{PP}}  - \chi^{s}  \right).
\end{aligned}
\end{equation}
The susceptibility in the triplet channel $\chi^{t}=\chi^{\text{PP}}_{\uparrow\uparrow\downarrow\downarrow}+\chi^{\text{PP}}_{\uparrow\downarrow\downarrow\uparrow} $ reads as
\begin{equation}
\label{eq:t-gen-susceptibility}
\begin{aligned}
\chi^{t}&=\chi^{(0),\text{PP}}_{\uparrow\uparrow\downarrow\downarrow} + \chi^{(0),\text{PP}}_{\uparrow\downarrow\downarrow\uparrow}\\
&-\frac{1}{2} \sum_{\left\{ \sigma_i \right\}}\chi^{(0),\text{PP}}_{\uparrow\sigma_1\downarrow\sigma_2} \Gamma^{\text{PP}}_{\sigma_2 \sigma_4 \sigma_1 \sigma_3 }\left( \chi^{(0),\text{PP}}_{\sigma_3\uparrow \sigma_4 \downarrow  } + \chi^{\text{PP}}_{\sigma_3\uparrow \sigma_4 \downarrow  }  \right) \\
&-\frac{1}{2} \sum_{\left\{ \sigma_i \right\}}\chi^{(0),\text{PP}}_{\uparrow\sigma_1 \downarrow \sigma_2} \Gamma^{\text{PP}}_{\sigma_2 \sigma_4 \sigma_1 \sigma_3 }\left( \chi^{(0),\text{PP}}_{\sigma_3\downarrow \sigma_4 \uparrow   } + \chi^{\text{PP}}_{\sigma_3\downarrow \sigma_4 \uparrow   }  \right)\\ 
&=\chi^{(0),\text{PP}}-\frac{1}{2}  \chi^{(0),\text{PP}} \Gamma^{t} \left(  \chi^{(0),\text{PP}}  + \chi^{t}  \right).
\end{aligned}
\end{equation}
Similarly to Eqs.~(\ref{eq:m-F-vertex}) and (\ref{eq:d-F-vertex}), we find the following expressions for the full vertices in these channels:
\begin{equation}
\label{eq:s-F-vertex}
\begin{aligned}
F^{s}&=\Gamma^{s}-\frac{1}{2} \Gamma^{s} \chi^{0,\text{PP}}F^{s},\\
\end{aligned}
\end{equation}
\begin{equation}
\label{eq:t-F-vertex}
\begin{aligned}
F^{t}&=\Gamma^{t}-\frac{1}{2} \Gamma^{t} \chi^{0,\text{PP}}F^{t}.\\
\end{aligned}
\end{equation}
Hence, the generalized susceptibilities read as:
\begin{equation}
\label{eq:gen-chi-s}
\begin{aligned}
\chi^{s}&=-\left[ \mathbb{I} -\frac{1}{2}  \chi^{(0),\text{PP}} \Gamma^{s} \right]^{-1} \\
&\times \left[ \chi^{(0),\text{PP}}+\frac{1}{2}  \chi^{(0),\text{PP}} \Gamma^{s}  \chi^{(0),\text{PP}}  \right],\\
\end{aligned}
\end{equation}
\begin{equation}
\label{eq:gen-chi-t}
\begin{aligned}
\chi^{t}&=\left[ \mathbb{I} + \frac{1}{2}  \chi^{(0),\text{PP}} \Gamma^{t} \right]^{-1} \\
&\times \left[ \chi^{(0),\text{PP}}-\frac{1}{2}  \chi^{(0),\text{PP}} \Gamma^{t}  \chi^{(0),\text{PP}}  \right],
\end{aligned}
\end{equation}
with the following approximated vertices
\begin{equation}
\Gamma^{s}\simeq
\left(\begin{array}{cccc}
2V_0&0&0&0\\
0&V_1&V_1&0\\
0&V_1&V_1&0\\
0&0&0&2V_0
\end{array}\right),
\end{equation}
\begin{equation}
\Gamma^{t}\simeq
\left(\begin{array}{cccc}
0&0&0&0\\
0&-V_1&V_1&0\\
0&V_1&-V_1&0\\
0&0&0&0
\end{array}\right).
\end{equation}
We note that the ordering in the matrix form by the spin indices for corresponding PH, vPH, and PP channels is determined using the crossing relations~\cite{PhysRevB.105.155101,PhysRevB.86.125114}.

For the singlet and triplet channels, a phase transition occurs when the eigenvalues of the equations: $ \frac{1}{2}\chi^{(0),\text{PP}}\Gamma^{s} \phi =\lambda \phi$ and $-\frac{1}{2} \chi^{(0),\text{PP}}\Gamma^{t} \phi =\lambda \phi$, equal to one. Therefore, the intravalley phase transition in the singlet channel is determined by the condition
\begin{equation}
\label{eq:rpa-s-channel-solution}
V_0 \chi_{\tau,\tau}^{(0),\text{PP}}-1
=0.\\
\end{equation}
Respectively, the intervalley phase transitions in the singlet and triplet channels is governed by the condition
\begin{equation}
\label{eq:rpa-t-channel-solution}
V_1 \chi_{\tau,-\tau}^{(0),\text{PP}} -1=0.
\end{equation}
Since Eqs.~(\ref{eq:intra-pp-susceptibility}) and (\ref{eq:inter-pp-susceptibility}) yield $\chi_{\tau,\tau'}^{(0),\text{PP}} <0$, there are no solutions, hence no phase transitions occur in the PP channel in the random phase approximation.

From Eq.~(\ref{eq:PP-0-def}), the total momentum transfer $\mathbf{Q}$ of the intravalley susceptibility $\chi_{\tau,\tau}^{(0),\text{PP}}$ is nonzero since $\mathbf{Q}=\left( \mathbf{k} + \mathbf{K}_{\tau}\right)+\left( -\mathbf{k} + \mathbf{K}_{\tau}\right)=2\mathbf{K}_{\tau}$. Therefore, it is expected that $\chi_{\tau,\tau}^{\text{PP}}$ corresponds to the pair density wave order. The total momentum transfer $\mathbf{Q}$ of the intervalley susceptibility $\chi_{\tau,-\tau}^{(0),\text{PP}}$ is zero since $\mathbf{Q}=\left( \mathbf{k} + \mathbf{K}_{\tau}\right)+\left(- \mathbf{k} + \mathbf{K}_{-\tau}\right)=\mathbf{0}$. Hence, it indeed corresponds to the Cooper instability.

\subsection{Phase diagrams and estimates of critical temperatures}
\label{sunsec:est}

To determine which phase dominates in the selected domain, we compare the corresponding eigenvalues of the matrices $-\chi^{(0),\text{PH}}\Gamma^{d} $ and $ \chi^{(0),\text{PH}}\Gamma^{m} $. In the random phase approximation, the eigenvalues $\lambda^{\text{Stoner}\,(m)}$, $\lambda^{\text{Stoner}\,(d)}$, $\lambda^{\text{IVC}}$ are linear in $\tilde{\chi}$. Since $\tilde{\chi}$ does not depend on the interaction parameters $V_0$ and $V_1$, only a single phase (either Stoner instability in the magnetic channel, Stoner instability in the density channel, or IVC phase) can emerge within the RPA at zero displacement field $U$. However, varying the ratio between the interaction parameters $V_0$ and $V_1$ may lead to the emergence of different correlated phases. From Eqs.~(\ref{eq:rpa-t-st-m}), (\ref{eq:rpa-t-ivc-m}), and (\ref{eq:rpa-t-st-d}), the general scaling law for the critical temperatures is the same for all phases for $SU(2) \times SU(2)$-symmetric interaction. Therefore, it is sufficient to present numerical results for the $SU(4)$-symmetric case and study which phases emerges depending on the ratio between $V_0$ and $V_1$.

For $SU(4)$-symmetric interaction, the eigenvalue $ \lambda^{\text{IVC}}>  \lambda^{\text{Stoner}}$ in the region where correlated phases emerge, therefore, there is no Stoner instability in RPA. In the general case of $SU(2) \times SU(2)$-symmetric interaction, if $V_1 \chi_{\tau,-\tau}^{(0),\text{PH}} > V_0 \chi_{\tau,\tau}^{(0),\text{PH}}$ and $V_1 \chi_{\tau,-\tau}^{(0),\text{PH}} > \left( 2V_1-V_0 \right) \chi_{\tau,\tau}^{(0),\text{PH}}$, the IVC phase dominates in both channels, and there is no Stoner instability. In our model, this leads to the following inequality for the interaction parameters:
\begin{equation}
\label{eq:rpa-ratio-ivc}
\frac{N-3}{N-2} V_1< V_0 < \frac{N-1}{N-2} V_1.
\end{equation}

For the Stoner instability to emerge, the following inequality has to be satisfied in the magnetic channel:
\begin{equation}
\label{eq:rpa-ratio-st-m}
V_0 > \frac{N-1}{N-2} V_1,
\end{equation}
Respectively, in the density channel, we have
\begin{equation}
\label{eq:rpa-ratio-st-d}
V_0 < \frac{N-3}{N-2} V_1.
\end{equation}
Note that similar conditions for the interaction parameters  for Stoner instability in the magnetic and density in bilayer graphene were proposed in Ref.~\onlinecite{Mayrhofer2025}. 

From Eqs.~(\ref{eq:rpa-t-st-m}), (\ref{eq:rpa-t-ivc-m}), and (\ref{eq:rpa-t-st-d}), we observe that the critical temperature $T_{c}$ does not grow infinitely as the number of layers increases. Indeed, for $T_c$ in all channels for $SU(4)$-symmetric interaction, when $V_0=V_1=V$, we obtain
\begin{equation}
\begin{aligned}
T_{c,\,\infty}&= \lim_{N \rightarrow \infty}  V^{\frac{N}{N-2}} \gamma_1^{\frac{2-2N}{2-N}} v_F^{\frac{2N}{2-N}} \\
&\times \lim_{N \rightarrow \infty} \left[\Gamma \left(\frac{2}{N}\right) \frac{1-2^{2-\frac{2}{N}}}{\pi N} \zeta\left(\frac{2}{N}-1\right) \right]^{\frac{N}{N-2} } \\
&= \frac{V}{8 \pi} \frac{\gamma_1^2}{v_F^2} .
\end{aligned}
\end{equation}

\begin{figure}[b]
\includegraphics[width=.45\textwidth]{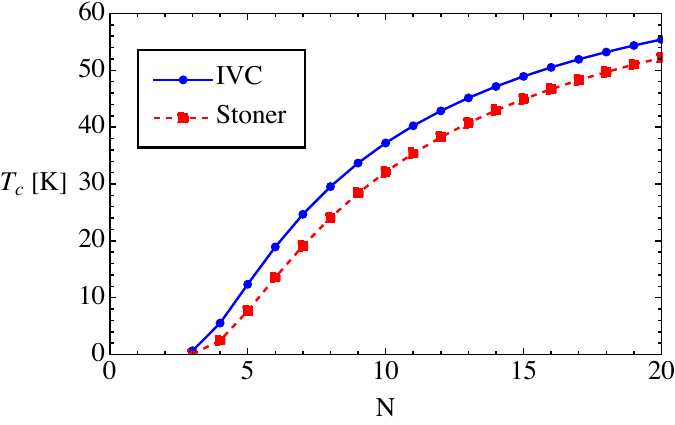}
\caption{Critical temperatures $T_c$ for the IVC and Stoner phases as a function of $N$ for $SU(4)$-symmetric interaction at $\mu=0$. The results for the Stoner phases given by Eqs.~(\ref{eq:rpa-t-st-m}) and (\ref{eq:rpa-t-st-d}) are identical in both channels because of $V_0=V_1$.}
\label{fig:beta_c_T}
\end{figure}

For a more quantitative analysis, we also calculate the values of $T_c $ in the $SU(4)$-symmetric case, for the parameters given in Table~\ref{tab:table-num}. The numerical values for the critical temperatures are highly dependent on the value of the intravalley interaction parameter in rhombohedral graphene. We expect the intravalley interaction parameter to be within 
\begin{equation}
9.3\,A_{uc}\,\, \mbox{eV}^{-1}<V_0<17.0\,A_{uc}\,\, \mbox{eV}^{-1} ,
\end{equation}
where the upper and lower limits correspond to the bare and constrained random phase approximation (cRPA) interaction parameter calculated in Ref.~\onlinecite{PhysRevLett.106.236805}. Therefore, we set $V_0=10\,A_{uc}\,\, \mbox{eV}^{-1}$.
\begin{table}[t]
\caption{\label{tab:table-num} Numerical values for the hopping parameters $\gamma_0$, $\gamma_1$, lattice constant $a$, area of a unit cell $A_{uc}=\sqrt{3} a^2/2$, Fermi velocity $v_F=\sqrt{3} a \gamma_0 /2$, and intravalley repulsive interaction parameter $V_0$. All parameters are presented in conventional and natural units (NU) and taken from Refs.~\onlinecite{McCann_2013,PhysRevB.80.165406, PhysRevB.110.045427,  PhysRevLett.132.186401,PhysRevLett.106.236805}.}
\begin{ruledtabular}
\begin{tabular}{lll}
Parameter & Numerical value  & Numerical value (NU) \\ \hline
 $\gamma_0$ & $3.16 \,\mbox{eV}$& $3.16 \,\mbox{eV}$ \\ 
$\gamma_1$  & $0.381 \,\mbox{eV}$ & $0.381 \,\mbox{eV}$ \\ 
 $a$ & $2.46 \,{\mbox{\r{A}}}$& $1.25 \cdot 10^{-3}\,\mbox{eV}^{-1}$\\ 
$v_F$ & $1.024\cdot 10^{6}\, \mbox{m/s}$ & $3.42 \cdot 10^{-3}\,\mbox{eV}^{0}$ \\ 
$A_{uc}$ & $5.24 \,{\mbox{\r{A}}}^2$ & $1.35 \cdot 10^{-6}\,\mbox{eV}^{-2}$ \\ 
$V_0$  &  $10 A_{uc}\,\, \mbox{eV} \cdot {\mbox{\r{A}}}^2$ &  $10 A_{uc}\,\, \mbox{eV}^{-1}$ \\ 
\end{tabular}
\end{ruledtabular}
\end{table}
The dependence of the critical temperatures $T_{c}[\mbox{K}]$ for the IVC (blue dots) and Stoner (red squares), given by Eqs.~(\ref{eq:rpa-t-st-m}), (\ref{eq:rpa-t-ivc-m}), and (\ref{eq:rpa-t-st-d}), on the number of layers $N$ is presented in Fig.~\ref{fig:beta_c_T}.
For these values for the physical parameters, we obtain the following result:
\begin{equation}
T_{c,\,\infty}= \frac{V}{8 \pi} \frac{\gamma_1^2}{v_F^2}\approx 77\,\mbox{K} .
\end{equation}

\begin{figure}[t]
\includegraphics[width=.45\textwidth]{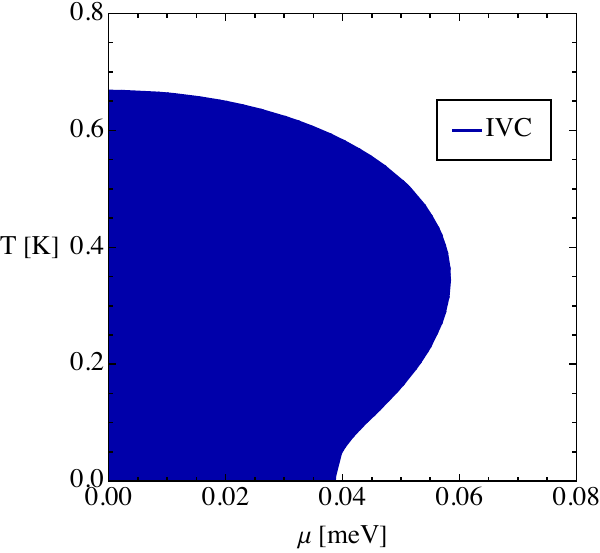}
\caption{Phase diagram $(\mu, T)$ for the IVC phase in the $SU(4)$-symmetric case for $N=3$ in the random phase approximation.}
\label{fig:phase_diagram-rpa}
\end{figure}

For nontrivial $\mu$, we present the phase diagram in coordinates $(\mu,T)$ by solving Eqs.~(\ref{M-phase-tr-Stoner}), (\ref{M-phase-tr-IVC}) and (\ref{D-phase-tr-Stoner}) in Fig.~\ref{fig:phase_diagram-rpa}. As stated in the beginning of Sec.~\ref{sunsec:est}, the IVC phase dominates in the $SU(4)$-symmetric case.
The critical temperature decreases as $\mu$ grows from zero. For the $SU(2)\times SU(2)$-symmetric case, the phase diagram would be similar to the one presented in Fig.~\ref{fig:phase_diagram-rpa}, albeit for a different phase, depending on the ratio between the interaction parameters $V_0$ and $V_1$, given in Eqs.~(\ref{eq:rpa-ratio-ivc}), (\ref{eq:rpa-ratio-st-m}), and (\ref{eq:rpa-ratio-st-d}).

Here, we observe that for some $\mu > \mu_0$ there exist two solutions for the critical temperature $T_{c}$, where, for a phase with corresponding eigenvalue $\lambda$, $\mu_0 $ is determined by the condition $\lambda(T=0,\mu_0)=1$, which gives
\begin{equation}
\mu_0 = \left[ \frac{\lambda}{\tilde{\chi}} \frac{1}{2\pi N \left( g_N \right)^{\frac{2}{N}}} \right]^{\frac{N}{N-2}}.
\end{equation}
This follows from the non-monotonic behavior of $\tilde{\chi}$ as a function of temperature at fixed $\mu$. Consequently, at fixed $\mu$, the maximum of the function $\tilde{\chi}(T,\mu)$ shifts away from $T=0$. 
After taking the partial derivative of $\tilde{\chi}(T,\mu)$ over $T$ at constant $\mu$, we obtain that the maximum is at the point $T= t \mu$, where $t $ is a solution of the following equation
\begin{equation}
\label{eq:sol1-axillary-1}
0=  1- \frac{2}{N}+ \frac{1}{t} \frac{\sum_{\lambda=\pm} \lambda \, \text{Li}_{\frac{2}{N}-2}\left( -e^{\lambda / t}\right)}{\sum_{\lambda=\pm}\text{Li}_{\frac{2}{N}-1}\left( -e^{\lambda / t}\right)} . 
\end{equation}
Eq.~(\ref{eq:sol1-axillary-1}) has one solution for $t$ for all $N \ge 3$. The dependence $t$ on $N$ is presented in Fig.~\ref{fig:t-n-axillary} in Appendix~\ref{app:Kubo}. As the number of layers increases, the function $t(N)$ approaches the solution of the equation
\begin{equation}
\label{eq:sol1-axillary-2}
\tanh\frac{1}{2 t} =t,
\end{equation}
which is obtained from Eq.~(\ref{eq:sol1-axillary-1}) in the limit $N\rightarrow \infty$. 

This is related to a similar non-monotonic behavior of the temperature-broadened density of states. For the density of states at $T=0$, we obtain the following expression:
\begin{equation}
\nu(\varepsilon)=\text{Tr} \left[ \delta \left( H_{\text{eff}} - \varepsilon\,\mathbb{I}_2  \right)\right]=  \frac{ 1}{2 \pi N \left(g_{N}\right)^\frac{2}{N}} \frac{1}{|\varepsilon|^{1-\frac{2}{N}  }} .
\end{equation}
Then, the temperature broadened density of states reads as
\begin{equation}
\nu_{T}(\mu)= 
\int_{-\infty}^{+\infty} d\omega \left( -\frac{\partial f(\omega)}{\partial \omega}\right) \nu(\omega) = \tilde{\chi}.
\end{equation}
It has the same behavior as the intravalley particle-hole susceptibility $\tilde{\chi}$. This coincidence between the susceptibility and the temperature-broadened density of states is not a general feature. Nevertheless, in our case, the density of states $\nu_{T}(\mu)$ increases with the temperature in the region $T \in \left[ 0 ,t \mu \right]$.

We demonstrate the non-monotonic behavior of $\tilde{\chi}$ with the temperature for several values of $\mu$ in Fig.~\ref{fig:tilde-chi-t}(a) of Appendix~\ref{app:Kubo}, using the two-band effective low-energy model given by Eq.~(\ref{eq:eff-hamiltonian}). 
This behavior can also be observed in the susceptibilities even for full models. Employing the full model for rhombohedral trilayer graphene with realistic hopping parameters, given by Eq.~(\ref{eq:RTG-DFT-model}), we calculate the intra- and intervalley susceptibilities. As shown in Fig.~\ref{fig:tilde-chi-t}(b) of Appendix~\ref{app:Kubo}, these susceptibilities also have such a distinctive peak at $\mu \neq 0$. Such reentrant behavior was also obtained in Ref~\onlinecite{ljrl-dg2n}.

\section{Parquet approximation}
\label{sec:parquet}

In the parquet approximation, we go deeper on level of the two-particle diagrammatics~\cite{PhysRevB.86.125114,RevModPhys.90.025003}, by approximating the channel-irreducible vertices $\Gamma$ in Eq.~(\ref{eq:app_ir_vertex}) as stated in Sec.~\ref{sec:susceptibility-gen}. The limit of applicability of the parquet approximation is determined by the solutions obtained within the RPA. Indeed, as we will show later, the poles in the approximated vertices $\Gamma^{s}$, $\Gamma^{t}$, $\Gamma^{m}$, $\Gamma^{d}$ will correspond to the solutions given by Eqs.~(\ref{M-phase-tr-Stoner}), (\ref{M-phase-tr-IVC}), and (\ref{D-phase-tr-Stoner}).

\subsection{$SU(4)$-symmetric interaction}

We begin by determining the constrains for the solutions in the parquet approximation for the $SU(4)$-symmetric interaction. These constrains for the $\tilde{\chi}$ are obtained from the following conditions:
\begin{equation}
\label{eq:limit-applicability-su4}
\begin{aligned}
1-V \chi_{\tau,\tau}^{(0),\text{PH}}&\ge 0,\\
1-V \chi_{\tau,-\tau}^{(0),\text{PH}}&\ge 0.\\
\end{aligned}
\end{equation}
Therefore, by choosing the strongest inequality, the region of the parquet approximation applicability in $(\beta,\mu)$ coordinates is determined by the condition; see Eq.~(\ref{eq:susceptibilities-via-tilde}):
\begin{equation}
\begin{aligned}
\frac{N-1}{N-2}V \tilde{\chi}(\beta,\mu) & \le 1,
\end{aligned}
\end{equation}
In Fig.~\ref{fig:beta_c_T_su4}, these solutions at $\mu=0$ are shown by the black line. Hence, every solution for a critical temperature lying below this line corresponds to a regime where the parquet approximation is not applicable.

In Sec.~\ref{sec:rpa}, we have shown in Eqs.~(\ref{eq:rpa-s-channel-solution}) and (\ref{eq:rpa-t-channel-solution}) that no phase transitions occur in the PP channel in the RPA. However, some solutions may exist in the PP channel within the parquet approximation. 

\subsubsection{Singlet and triplet channels}
\label{subsec-st-su4}

From Eq.~(\ref{eq:app_ir_vertex}), we obtain the following expressions for the channel-irreducible vertices $\Gamma^{s}$ and $\Gamma^{t}$ (see Appendix~\ref{app:explicit-vertices} for details):
\begin{equation}
\label{eq:irr-vertex-s}
\begin{aligned}
\Gamma^{s} \left( k , k^{\prime},q\right)&=\Lambda^{s} + \left[ \frac{1}{2} \Phi^{d}- \frac{3}{2} \Phi^{m} \right]  \left( k , k^{\prime},q-k - k^{\prime}\right)\\
&+ \left[ \frac{1}{2} \Phi^{d}- \frac{3}{2} \Phi^{m} \right] \left( k ,q- k^{\prime},-k + k^{\prime}\right), \\
\end{aligned}
\end{equation}
\begin{equation}
\label{eq:irr-vertex-t}
\begin{aligned}
\Gamma^{t} \left( k , k^{\prime},q\right)&=\Lambda^{t} + \left[ \frac{1}{2} \Phi^{d}+\frac{1}{2} \Phi^{m} \right]  \left( k , k^{\prime},q-k - k^{\prime}\right)\\
&- \left[ \frac{1}{2} \Phi^{d}+ \frac{1}{2} \Phi^{m} \right]\left( k ,q- k^{\prime},-k + k^{\prime}\right) . \\
\end{aligned}
\end{equation}
The ordering in the matrix form by the spin indices for corresponding PH, vPH, and PP channels is done using crossing relations, as in Sec.~\ref{sec:rpa}.

We use the approximated vertices with $V_0=V_1=V$, given by Eqs.~(\ref{eq:app-irr-vertex-s-flex}) and (\ref{eq:app-irr-vertex-t-flex}), in Eqs.~(\ref{eq:gen-chi-s}) and (\ref{eq:gen-chi-t}) and identify a phase transition when the eigenvalues of the following equations: $-\frac{1}{2} \chi^{(0),\text{PP}}\Gamma^{t} \phi =\lambda \phi$ and $ \frac{1}{2} \chi^{(0),\text{PP}}\Gamma^{s} \phi =\lambda \phi$, equal to one. 

In the singlet channel, we obtain the following eigenvalue for the pair-density wave (PDW):
\begin{equation}
\label{eq:flex-cooper-s}
\begin{aligned}
\lambda^{\text{PDW}}&=- V \chi_{\tau,\tau}^{(0),\text{PP}}\\
&-\frac{V \chi_{\tau,\tau}^{(0),\text{PP}}( 3 V \chi_{\tau,\tau}^{(0),\text{PH}} + 2 )}{ 3 V^2 (\chi_{\tau,\tau}^{(0),\text{PH}} )^2 -2 V \chi_{\tau,\tau}^{(0),\text{PH}} - 1}.\\
\end{aligned}
\end{equation}
The corresponding eigenvectors are $\phi=\left( 1,0,0,0 \right)^{T}$  and $\phi=\left( 0,0,0,1 \right)^{T}$. 
Equation $\lambda^{\text{PDW}} = 1$ has no real solutions for $\tilde{\chi}$. As will be discussed later, the $SU(4)$-spin symmetry should be reduced to $SU(2) \times SU(2)$ to obtain solutions for the pair-density wave.

\begin{figure}[t]
\includegraphics[width=.45\textwidth]{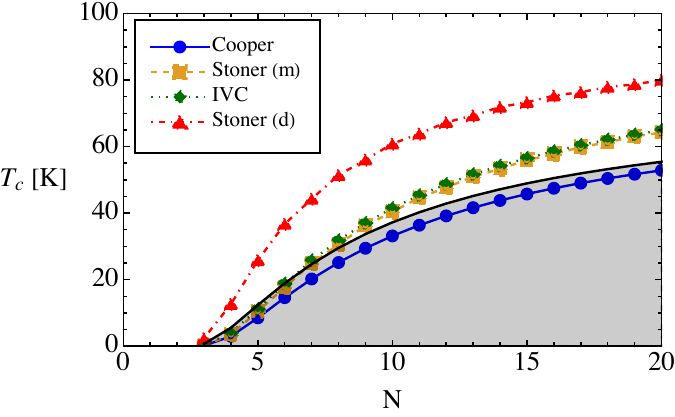}
\caption{Critical temperature $T_c$ for the correlated phases as a function of $N$ for $SU(4)$-symmetric interaction in the parquet approximation at $\mu=0$. The parquet approximation is not applicable below the black line in the gray region. In this region, the random phase approximation should be used.}
\label{fig:beta_c_T_su4}
\end{figure}

In the triplet and singlet channels, the eigenvalue for the intervalley transitions that corresponds to Cooper instability is given by
\begin{equation}
\label{eq:flex-cooper-t}
\begin{aligned}
\lambda^{\text{Cooper}}&=-V \chi_{\tau,-\tau}^{(0),\text{PP}}-\frac{V \chi_{\tau,-\tau}^{(0),\text{PP}}}{V \chi_{\tau,-\tau}^{(0),\text{PH}}-1}\\
&-\frac{V \chi_{\tau,-\tau}^{(0),\text{PP}}}{ 3 V^2 (\chi_{\tau,\tau}^{(0),\text{PH}} )^2 -2 V \chi_{\tau,\tau}^{(0),\text{PH}} - 1} . \\
\end{aligned}
\end{equation}
In this case, the leading eigenvector in the singlet channel is $\phi=\left( 0,1,1,0 \right)^{T}$, which corresponds to $s$-wave superconductivity, and the eigenvector in the triplet channel is $\phi=\left( 0,1,-1,0 \right)^{T}$, which corresponds to $f$-wave superconductivity~\cite{Hrhold2023}. Eq.~(\ref{eq:flex-cooper-t}) has a one solution for $\tilde{\chi}$ for each $N \ge 3$.  The dependence of the critical temperature $T_{c}$ for the Cooper instability on the number of layers at the charge neutrality point is presented in Fig.~\ref{fig:beta_c_T_su4} (blue dots) for the values of the parameters, given in Table~\ref{tab:table-num}. 

However, the solutions for $\tilde{\chi}$ of equation $\lambda^{\text{Cooper}}=1$ are beyond the range of applicability of the parquet approximation for all $N \ge 3$, since these solutions lie beyond the point where the channel-reducible vertices $\Phi$ are divergent. Hence, all solution for $T_c$ lie below the black line in Fig.~\ref{fig:beta_c_T_su4}, which we show explicitly for illustration purposes only. Therefore, Cooper instability cannot be described within our approximations. 

\subsubsection{Magnetic and density channels}
\label{subsec-md-su4}

In the magnetic and density channels, the channel irreducible vertices are given by (see Appendix~\ref{app:explicit-vertices} for details):
\begin{equation}
\label{eq:irr-vertex-m}
\begin{aligned}
\Gamma^{m}\left( k , k^{\prime},q\right)&=\Lambda^{m} - \left[ \frac{1}{2} \Phi^{t} - \frac{1}{2} \Phi^{s} \right] \left( k , k^{\prime},q+ k + k^{\prime}\right) \\
&+ \left[ \frac{1}{2} \Phi^{d}- \frac{1}{2} \Phi^{m}\right] \left( k , q+k,- k + k^{\prime}\right)  , \\
\end{aligned}
\end{equation}
\begin{equation}
\label{eq:irr-vertex-d}
\begin{aligned}
\Gamma^{d}\left( k , k^{\prime},q\right)&=\Lambda^{d} + \left[ \frac{3}{2} \Phi^{t} + \frac{1}{2} \Phi^{s} \right] \left( k , k^{\prime},q+ k + k^{\prime}\right) \\
&- \left[ \frac{1}{2} \Phi^{d}+ \frac{3}{2} \Phi^{m}\right] \left( k , q+k,- k + k^{\prime}\right) .  \\
\end{aligned}
\end{equation}
Using the vertices given by Eqs.~(\ref{eq:app-irr-vertex-m-flex}) and (\ref{eq:app-irr-vertex-d-flex}), the eigenvalues for Stoner instability $\lambda^{\text{Stoner}}$ and the intervalley coherent phase $\lambda^{\text{IVC}}$ in the magnetic channel are given by
\begin{equation}
\label{M-phase-tr-Stoner-flex-1}
\begin{aligned}
\lambda^{\text{Stoner}\, (m)}&=-\frac{V^2 \chi_{\tau,\tau}^{(0),\text{PH}} \chi_{\tau,\tau}^{(0),\text{PP}}}{V \chi_{\tau,\tau}^{(0),\text{PP}}-1}\\
&- \frac{V \chi_{\tau,\tau}^{(0),\text{PH}}  }{ 3 V^2 (\chi_{\tau,\tau}^{(0),\text{PH}} )^2 -2 V \chi_{\tau,\tau}^{(0),\text{PH}} - 1}, \\
\end{aligned}
\end{equation}
\begin{equation}
\label{M-phase-tr-IVC-flex}
\begin{aligned}
\lambda^{\text{IVC}}&=-\frac{V^2 \chi_{\tau,-\tau}^{(0),\text{PH}} \chi_{\tau,-\tau}^{(0),\text{PP}}}{V \chi_{\tau,-\tau}^{(0),\text{PP}}-1} \\
&-\frac{V \chi_{\tau,-\tau}^{(0),\text{PH}} }{ 3 V^2 (\chi_{\tau,\tau}^{(0),\text{PH}} )^2 -2 V \chi_{\tau,\tau}^{(0),\text{PH}} - 1} , \\ 
\end{aligned}
\end{equation}
respectively. Eigenvalue $\lambda^{\text{Stoner}\, (m)}$ corresponds to the eigenvectors $\phi=\left(1,0,0,0\right)^{T}$ and $\phi=\left(0,0,0,1\right)^{T}$. For the intervalley coherence state, the eigenvectors are $\phi=\left(0,1,0,0\right)^{T}$ and $\phi=\left(0,0,1,0\right)^{T}$. 

After choosing the solution that corresponds to the highest temperature, we obtain numerical results for the dependence of the critical temperatures $T_{c}$ for the Stoner (yellow squares) and the IVC (green diamonds) phases on $N$, which is presented in Fig.~\ref{fig:beta_c_T_su4} for the parameter values given in Table~\ref{tab:table-num}. 
Here, we observe that for these values parameters only for $N \ge 7$, the solutions for the critical temperatures are within the parquet approximation. In the limit $N \rightarrow \infty$, we obtain the following result:
\begin{equation}
\label{eq:T-limit-magnetic-parquet}
T_{c,\,\infty}^{\text{Stoner}\,(m)}=T_{c,\,\infty}^{\text{IVC}}\simeq\frac{V}{20.79} \frac{\gamma_1^2}{v_F^2}.
\end{equation}
Using Table~\ref{tab:table-num}, we estimate $T_{c,\,\infty}^{\text{Stoner}\,(m)}= T_{c,\,\infty}^{\text{IVC}\,(m)} \approx 93\,\mbox{K}$.

In the density channel, the eigenvalues corresponding to the Stoner instability are:
\begin{equation}
\label{D-phase-tr-Stoner-flex-1}
\begin{aligned}
\lambda^{(\mu)}&=\frac{V \chi_{\tau,\tau}^{(0),\text{PH}}(1+ 6 V \chi_{\tau,\tau}^{(0),\text{PH}} )  }{ 3 V^2 (\chi_{\tau,\tau}^{(0),\text{PH}} )^2 -2 V \chi_{\tau,\tau}^{(0),\text{PH}} - 1}\\
& +  \frac{2V \chi_{\tau,\tau}^{(0),\text{PH}}}{ V \chi_{\tau,-\tau}^{(0),\text{PH}} -1 } + \frac{V \chi_{\tau,\tau}^{(0),\text{PH}}   }{  V \chi_{\tau,\tau}^{(0),\text{PP}} -1 } \\
&+\frac{2 V \chi_{\tau,\tau}^{(0),\text{PH}}   }{ V \chi_{\tau,-\tau}^{(0),\text{PP}} -1 }  +3 V \chi_{\tau,\tau}^{(0),\text{PH}},
\end{aligned}
\end{equation}
\begin{equation}
\label{D-phase-tr-Stoner-flex-2}
\begin{aligned}
\lambda^{\text{Stoner}\, (d)}&= \frac{(V \chi_{\tau,\tau}^{(0),\text{PH}})^2 (4+ 3 V \chi_{\tau,\tau}^{(0),\text{PH}} )  }{ 3 V^2 (\chi_{\tau,\tau}^{(0),\text{PH}} )^2 -2 V \chi_{\tau,\tau}^{(0),\text{PH}} - 1}\\
& -  \frac{2V \chi_{\tau,\tau}^{(0),\text{PH}}}{ V \chi_{\tau,-\tau}^{(0),\text{PH}} - 1} + \frac{V \chi_{\tau,\tau}^{(0),\text{PH}}   }{  V \chi_{\tau,\tau}^{(0),\text{PP}} -1 } \\
&-\frac{2 V \chi_{\tau,\tau}^{(0),\text{PH}}   }{ V \chi_{\tau,-\tau}^{(0),\text{PP}} -1 }-2 V \chi_{\tau,\tau}^{(0),\text{PH}}.
\end{aligned}
\end{equation}
For the IVC in the density channel, the eigenvalue
coincides with Eq.~(\ref{M-phase-tr-IVC-flex}). For the intervalley coherence state, the eigenvectors are $\phi=\left(0,1,0,0\right)^{T}$ and $\phi=\left(0,0,1,0\right)^{T}$. Eq.~(\ref{D-phase-tr-Stoner-flex-2}) corresponds to the eigenvector $\phi=\left(1,0,0,-1\right)^{T}$, therefore this equation corresponds to the valley polarization. 

Eq.~(\ref{D-phase-tr-Stoner-flex-1}) corresponds to the eigenvector $\phi=\left(1,0,0,1\right)^{T}$, therefore this equation corresponds to the chemical potential instability, i.e. $M_{\Phi}(\mathbf{k},\mathbf{k}^{\prime})=s_{0} \otimes \tau_{0} \, \delta(\mathbf{k}-\mathbf{k}^{\prime})$. The solutions for the $\lambda^{(\mu)}$ are always beyond the limits of applicability of the parquet approximation. Moreover, this instability is considered trivial, as the associated order parameter—the particle number—does not break any symmetry.

In Fig.~\ref{fig:beta_c_T_su4}, we present the numerical results for the dependence of the critical temperatures $T_{c}$ for the Stoner (red triangles) and the IVC (green diamonds) phases on $N$. In the limit $N \rightarrow \infty$, we obtain
\begin{equation}
\label{eq:T-limit-density-parquet}
T_{c,\,\infty}^{\text{Stoner}\,(d)}=T_{c,\,\infty}^{\text{IVC}}\simeq\frac{V}{20.79} \frac{\gamma_1^2}{v_F^2},\\
\end{equation}
which coincides with Eq.~(\ref{eq:T-limit-magnetic-parquet}) for the magnetic channel.

\subsubsection{Phase diagrams}
\label{subsec-phase-su4}

Summarizing the results in Secs.~\ref{subsec-st-su4} and \ref{subsec-md-su4}, we may see that the dependence of the critical temperature for all instabilities on the number of layers has the following form
\begin{equation}
\label{eq:temperature-scaling-su4}
\begin{aligned}
T_c(N)&=V^{\frac{N}{N-2}} \gamma_1^{\frac{2-2N}{2-N}} v_F^{\frac{2N}{2-N}}  \\
&\times \left[\frac{1}{x_{N}}\Gamma \left(\frac{2}{N}\right) \frac{1-2^{2-\frac{2}{N}}}{\pi N} \zeta\left(\frac{2}{N}-1\right) \right]^{\frac{N}{N-2} }
\end{aligned}
\end{equation}
where $x_{N}$ is the smallest positive solution for $V \tilde{\chi}$ of the corresponding algebraic equation $\lambda=1$ within the region of applicability of the parquet approximation. This presents a general scaling law for the critical temperature with the number of layers in the R$N$G. 

\begin{figure}[t]
\includegraphics[width=.45\textwidth]{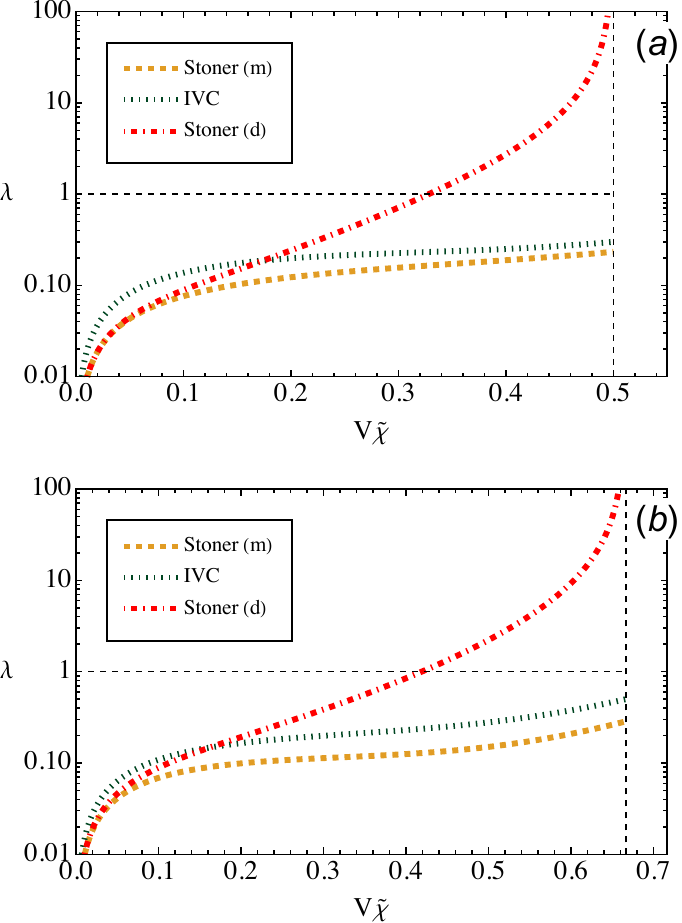}
\caption{Dependence of the eigenvalues on $V \tilde{\chi}$ for the Stoner instabilities and IVC phase in the magnetic and density channels in the case of $SU(4)$-symmetric interaction and (a) $N=3$, (b) $N=4$. The horizontal dashed line corresponds to $\lambda = 1$. The parquet approximation is applicable to the left of the vertical dashed line. }
\label{fig:phase_diagram-lambda-parquet}
\end{figure}

\begin{figure}[t]
\includegraphics[width=.45\textwidth]{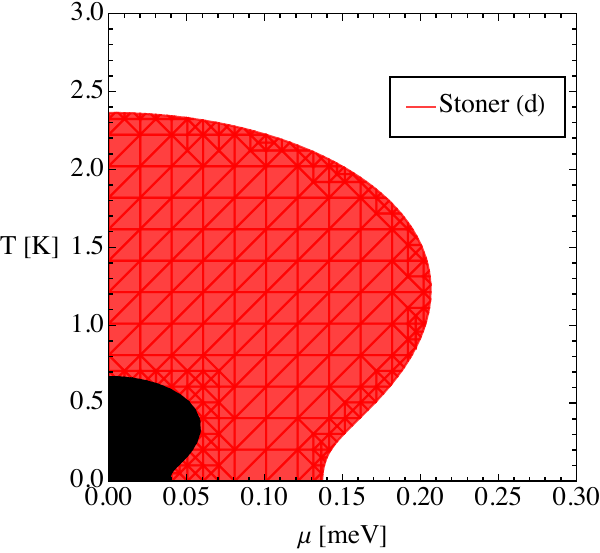}
\caption{Phase diagram for the Stoner instability in the density channel in the parquet approximation in $SU(4)$-symmetric case and $N=3$ in coordinates $(\mu, T)$. The parquet approximation is not applicable in the black region.}
\label{fig:phase_diagram-parquet}
\end{figure}

The phase diagram is constructed by comparing the eigenvalues with $\lambda \ge 1$ and choosing the region where the corresponding eigenvalue is the largest.  For $SU(4)$-symmetric interaction, we obtain that the Stoner instability in the density channel dominates for each $N \ge 3$. The corresponding region in $(\beta,\mu)$ coordinates is determined by the following inequality
\begin{equation}
x^{\text{Stoner}\, (d)}_{N}<V\tilde{\chi} \left( \beta, \mu \right) < \frac{N-2}{N-1} ,
\end{equation}
where $x^{\text{Stoner}\, (d)}_{N}$ is the smallest positive solution for $V \tilde{\chi}$ of Eq~(\ref{D-phase-tr-Stoner-flex-2}).
This dependence of eigenvalues on $V \tilde{\chi}(\beta,\mu)$ for $N=3,4$ is explicitly shown in Fig.~\ref{fig:phase_diagram-lambda-parquet}. 
The phase diagram in $(\mu,T)$ coordinates, corresponding to Fig.~\ref{fig:phase_diagram-lambda-parquet} (a), is presented in Fig.~\ref{fig:phase_diagram-parquet}. 

We obtain that the Stoner instability in the density channel dominates for $SU(4)$-symmetric interaction in the parquet approximation. There are no qualitative changes in the phase diagrams for the larger number of layers. Similarly to the results obtained in Sec.~\ref{sec:rpa}, we found that the critical temperature decreases for nonzero chemical potential. 

\subsection{$SU(2)\times SU(2)$-symmetric interaction}

The constraints for $\tilde{\chi}$ for $SU(2)\times SU(2)$-symmetric interaction are determined by the following inequalities [cf. with Eq.~(\ref{eq:limit-applicability-su4})]:
\begin{equation}
\label{eq:limit-applicability-su2}
\begin{aligned}
1-V_0 \chi_{\tau,\tau}^{(0),\text{PH}}& \ge 0,\\
1-\left(2V_1-V_0\right) \chi_{\tau,\tau}^{(0),\text{PH}} &\ge 0,\\
1-V_1 \chi_{\tau,-\tau}^{(0),\text{PH}} &\ge 0.\\
\end{aligned}
\end{equation}
Therefore, by choosing stronger inequality, the region of the parquet approximation applicability in $(\beta,\mu)$ coordinates are determined by the following inequality:
\begin{equation}
\label{eq:chi-l-def}
V_0 \tilde{\chi}(\beta,\mu) \le V_0 \tilde{\chi}^{(l)}=\min \left(\frac{N-2}{N-1}\frac{V_0}{V_1},\frac{V_0}{2 V_1-V_0},1\right).
\end{equation}

\subsubsection{Singlet and triplet channels}
\label{subsec-st-su2}

In the singlet channel, the eigenvalue for the intravalley transitions reads as follows:
\begin{equation}
\label{eq:flex-cooper-s-su2}
\begin{aligned}
\lambda^{\text{PDW}}&=\frac{1}{2}V_0 \chi_{\tau,\tau}^{(0),\text{PP}} -\frac{3}{2} \frac{V_0^2 \chi_{\tau, \tau}^{(0),\text{PP}} \chi_{\tau,\tau}^{(0),\text{PH}} }{V_0\chi_{\tau, \tau}^{(0),\text{PH}} -1}\\
&+ \frac{1}{2} \frac{V_0\chi_{\tau,\tau}^{(0),\text{PP}}+ \left(V_0^2-4V_1^2\right) \chi_{\tau,\tau}^{(0),\text{PP}} \chi_{\tau,\tau}^{(0),\text{PH}}}{(V_0 \chi_{\tau, \tau}^{(0),\text{PH}} +1 )^2-4 V_1^2 (\chi_{\tau, \tau}^{(0),\text{PH}})^2}.
\end{aligned}
\end{equation}

In the singlet and triplet channels, the eigenvalue for the intervalley transition is
\begin{equation}
\label{eq:flex-cooper-t-su2}
\begin{aligned}
\lambda^{\text{Cooper}}&=-V_1 \chi_{\tau,-\tau}^{(0),\text{PP}} - \frac{V_1 \chi_{\tau, -\tau}^{(0),\text{PP}} }{V_1\chi_{\tau, -\tau}^{(0),\text{PH}} -1}\\
&+ \frac{V_1\chi_{\tau,-\tau}^{(0),\text{PP}}}{(V_0 \chi_{\tau, \tau}^{(0),\text{PH}} +1 )^2-4 V_1^2 (\chi_{\tau, \tau}^{(0),\text{PH}})^2} .\\
\end{aligned}
\end{equation}

\begin{figure}[b]
\includegraphics[width=.45\textwidth]{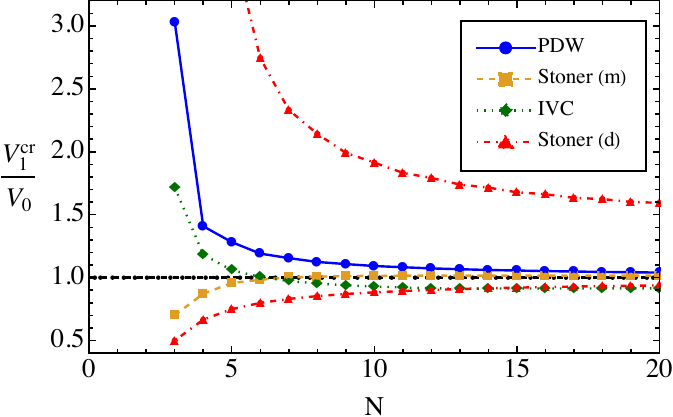}
\caption{Dependencies of the critical interaction parameters $V_1^{\textit{cr}}$ on the number of layers $N$ in units of $V_0$ for corresponding correlated phases. In all channels, $V_{1}^{\textit{cr}} \rightarrow V_0$ as $N \rightarrow \infty$. 
}
\label{fig:v1-crit}
\end{figure}

\begin{figure}[t]
\includegraphics[width=.45\textwidth]{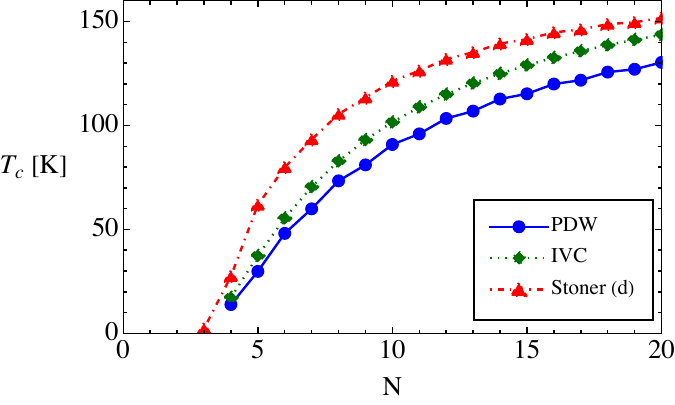}
\caption{Critical temperature $T_c$ for the correlated phases as a function of $N$ for $SU(2)\times SU(2)$-symmetric interaction in the parquet approximation for $V_1=3V_0/2$ at $\mu=0$. The solutions that are beyond the range of applicability of the parquet approximation are already excluded.}
\label{fig:beta_c_T-su2}
\end{figure}

Similarly to the Cooper instability in the case of the $SU(4)$-symmetric interaction, the solution for $T_{c}$, obtained from $\lambda^{\text{Cooper}}=1$, lies beyond the range of applicability of the parquet approximation. However, since in $SU(2)\times SU(2)$ case, the interaction parameters $V_1 \neq V_0$, we obtain that the solution for the pair-density wave phase exists for $V_1 > V_1^{\textit{cr}}$. The dependence of the critical interaction parameter $V_1^{\textit{cr}}$ on the number of layers $N$ for the PDW (blue dots) order is presented in Fig.~\ref{fig:v1-crit}. 
Here, we obtain that the PDW solution exists only if $V_1 > V_0$, which may be unlikely, since it is expected that the intervalley effective repulsive interaction is smaller than intravalley~\cite{PhysRevB.110.L201113}.
The dependence of the critical temperature $T_{c}$ on $N$ for the PDW (blue dots) phase is presented in Fig.~\ref{fig:beta_c_T-su2} for $V_1=3V_0/2$ and the numerical parameters, given in Table~\ref{tab:table-num}. 

\subsubsection{Magnetic and density channels}
\label{subsec-md-su2}

Using the vertices given by Eqs.~(\ref{eq:app-irr-vertex-m-flex}) and (\ref{eq:app-irr-vertex-d-flex}), the eigenvalues in the magnetic channel are given by
\begin{equation}
\label{M-phase-tr-Stoner-flex-1-su2}
\begin{aligned}
\lambda^{\text{Stoner}\, (m)}&=-\frac{V_0^2 \chi_{\tau,\tau}^{(0),\text{PH}} \chi_{\tau,\tau}^{(0),\text{PP}}}{V_0 \chi_{\tau,\tau}^{(0),\text{PP}}-1}+\frac{1}{1 - V_0  \chi_{\tau, \tau}^{(0),\text{PH}}}\\
&\times\frac{V_0 \chi_{\tau, \tau}^{(0),\text{PH}} +(V_0^2 -2V_1^2)  (\chi_{\tau, \tau}^{(0),\text{PH}})^2}{(V_0 \chi_{\tau, \tau}^{(0),\text{PH}} +1 )^2-4 V_1^2 (\chi_{\tau, \tau}^{(0),\text{PH}})^2} ,
\end{aligned}
\end{equation}
for the Stoner instability, and
\begin{equation}
\label{M-phase-tr-IVC-flex-su2}
\begin{aligned}
\lambda^{\text{IVC}}&=-\frac{V_1^2 \chi_{\tau,-\tau}^{(0),\text{PH}} \chi_{\tau,-\tau}^{(0),\text{PP}}}{V_1 \chi_{\tau,-\tau}^{(0),\text{PP}}-1} \\
&+\frac{V_1 \chi_{\tau,-\tau}^{(0),\text{PH}} }{(V_0 \chi_{\tau, \tau}^{(0),\text{PH}} +1 )^2-4 V_1^2 (\chi_{\tau, \tau}^{(0),\text{PH}})^2} , \\ 
\end{aligned}
\end{equation}
for the intervalley coherence phase. For the intervalley coherence state, the eigenvectors are $\phi=\left(0,1,0,0\right)^{T}$ and $\phi=\left(0,0,1,0\right)^{T}$. Eq.~(\ref{M-phase-tr-Stoner-flex-1-su2}) corresponds to the eigenvectors $\phi=\left(1,0,0,0\right)^{T}$ and $\phi=\left(0,0,0,1\right)^{T}$.

There is a solution for the IVC phase for $V_1>V_1^{\textit{cr}}$. Here, $V_1^{\textit{cr}}$ is the corresponding critical interaction parameter for the IVC phase. Similarly, we obtain a solution for the Stoner instability in the magnetic channel for $V_1$ smaller than corresponding critical parameter $V_1^{\textit{cr}}$. The dependencies of critical parameters $V_1^{\textit{cr}}$ on $N$ for the IVC (green diamonds) and Stoner instabilities (yellow squares) are presented in Fig.~\ref{fig:v1-crit}. 

\begin{figure}[t]
\includegraphics[width=.45\textwidth]{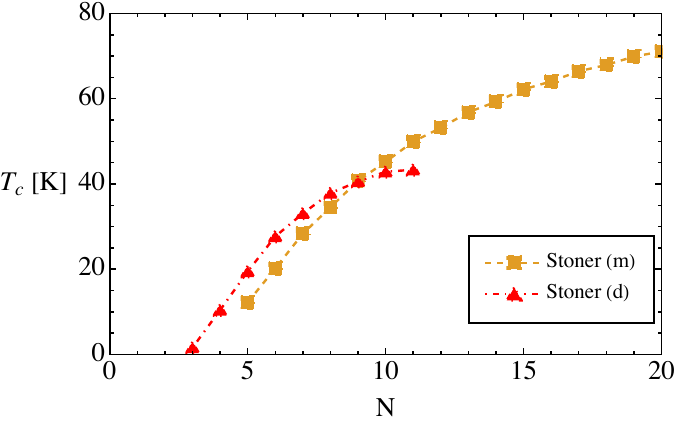}
\caption{Dependence of the critical temperature $T_c$ on $N \ge 3$ for $SU(2)\times SU(2)$-symmetric interaction in the parquet approximation for $V_1=0.9V_0$ at $\mu=0$. The solutions that are beyond the range of applicability of the parquet approximation are already excluded.}
\label{fig:beta_c_T-su2-2}
\end{figure}

In Fig.~\ref{fig:beta_c_T-su2}, we show the dependence of the critical temperature $T_{c}$ on $N$ for the IVC (green diamonds) phase and $V_1=3V_0/2$.
The dependence pf $T_c(N)$ on $N$ for the Stoner phase (yellow squares) for $V_1=0.9V_0$ is presented in Fig.~\ref{fig:beta_c_T-su2-2}. 

The eigenvalue for the Stoner instability in the density channel reads as follows:
\begin{equation}
\label{D-phase-tr-Stoner-flex-2-su2}
\begin{aligned}
\lambda^{\text{Stoner}\, (d)}&= \frac{1}{2} \frac{ V_0 \chi_{\tau,\tau}^{(0),\text{PH}} +  \left(V_0^2-4 V_1^2\right) (\chi_{\tau,\tau}^{(0),\text{PH}})^2 }{ (V_0 \chi_{\tau,\tau}^{(0),\text{PH}}+1)^2-4 V_1^2 (\chi_{\tau,\tau}^{(0),\text{PH}})^2}\\
&+\frac{3}{2} \frac{V_0 \chi_{\tau,\tau}^{(0),\text{PH}} }{V_0 \chi_{\tau,\tau}^{(0),\text{PH}} -1 }-\frac{2 V_1^2 \chi_{\tau,\tau}^{(0),\text{PH}} \chi_{\tau,-\tau}^{(0),\text{PH}}}{V_1 \chi_{\tau,-\tau}^{(0),\text{PH}}-1}\\
&+\frac{V_0^2 \chi_{\tau,\tau}^{(0),\text{PH}} \chi_{\tau,\tau}^{(0),\text{PP}}}{V_0 \chi_{\tau,\tau}^{(0),\text{PP}}-1}-\frac{2 V_1^2 \chi_{\tau,\tau}^{(0),\text{PH}} \chi_{\tau,-\tau}^{(0),\text{PP}} }{V_1 \chi_{\tau,-\tau}^{(0),\text{PP}}-1}\\
&+2 V_1 \chi_{\tau,\tau}^{(0),\text{PH}} .
\end{aligned}
\end{equation}
Eq.~(\ref{D-phase-tr-Stoner-flex-2-su2}) corresponds to the valley polarization since its corresponding eigenvector is $\phi=\left(1,0,0,-1\right)^{T}$. The eigenvalue $\lambda^{\text{IVC}}$ for the IVC in the density channel coincides with that given by Eq.~(\ref{M-phase-tr-IVC-flex-su2}). The eigenvalue corresponding to the eigenvector $\phi=\left(1,0,0,1\right)^{T}$, similarly to the $SU(4)$-symmetric case, leads to the solutions beyond the limit of applicability of the parquet approximation; therefore, these are dubbed unphysical.

The solution for Stoner instability in the density channel exists for $V_1^{\textit{cr},\,\text{l}}<V_1<V_1^{\textit{cr},\,\text{u}}$, where $V_1^{\textit{cr},\,\text{l}}$ and $V_1^{\textit{cr},\,\text{u}}$ correspond to the lower and upper bounds for $V_1$, presented with red triangles in Fig.~\ref{fig:v1-crit}. The numerical results for the critical temperature $T_{c}$ (red triangles) are presented in Fig.~\ref{fig:beta_c_T-su2} for $V_1=3V_0/2$ and in Fig.~\ref{fig:beta_c_T-su2-2} for $V_1=0.9V_0$.


\subsubsection{Phase diagrams}
\label{subsec-phase-su2}

\begin{figure}[t]
\includegraphics[width=.45\textwidth]{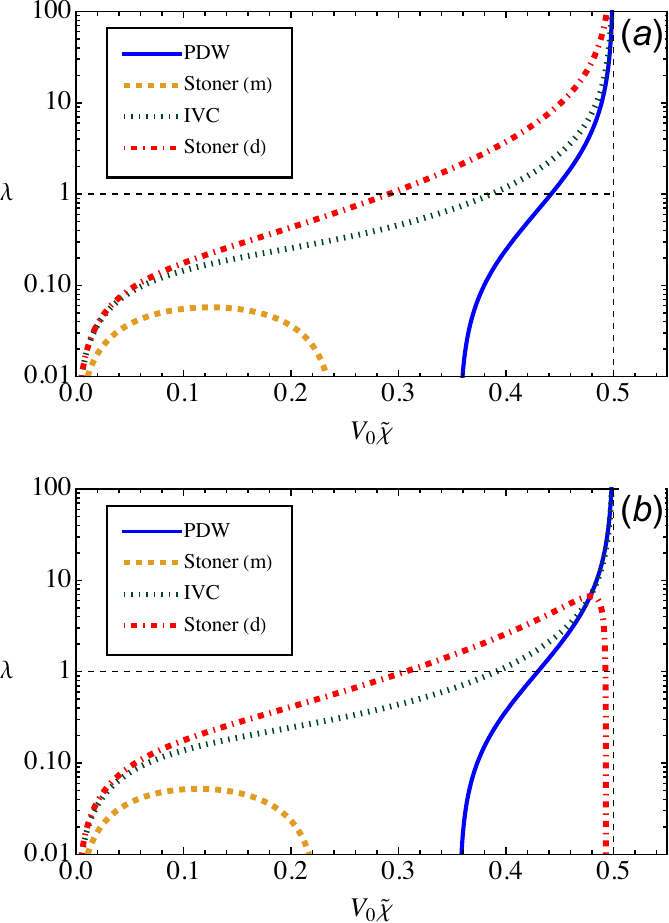}
\caption{Dependence of the eigenvalues for the correlated phases on $V_0 \tilde{\chi}$ for $SU(2)\times SU(2)$-symmetric interaction with $V_1=3V_0/2$ and (a) $N=5$, (b) $N=6$. The horizontal dashed line corresponds to $\lambda = 1$. The parquet approximation is applicable to the left of the vertical dashed line.
}
\label{fig:phase_diagram-lambda-parquet-su2}
\end{figure}

The dependence of eigenvalues on $V_0 \tilde{\chi}(\beta,\mu)$ for $N=5,6$ is explicitly shown in Fig.~\ref{fig:phase_diagram-lambda-parquet-su2} for $V_1=3V_0/2$.
In this case, the Stoner instability in the density channel (red dot-dashed line) dominates in the PH channel for $N <6$, while for $6 \le N \le 28$ there are regions with the Stoner and the IVC (green dotted line) phases, and for $N> 28$ there is only IVC in the PH channel. In the PP channel, the PDW phase (blue solid line) exists for $N \ge 4$.

\begin{figure}[t]
\includegraphics[width=.45\textwidth]{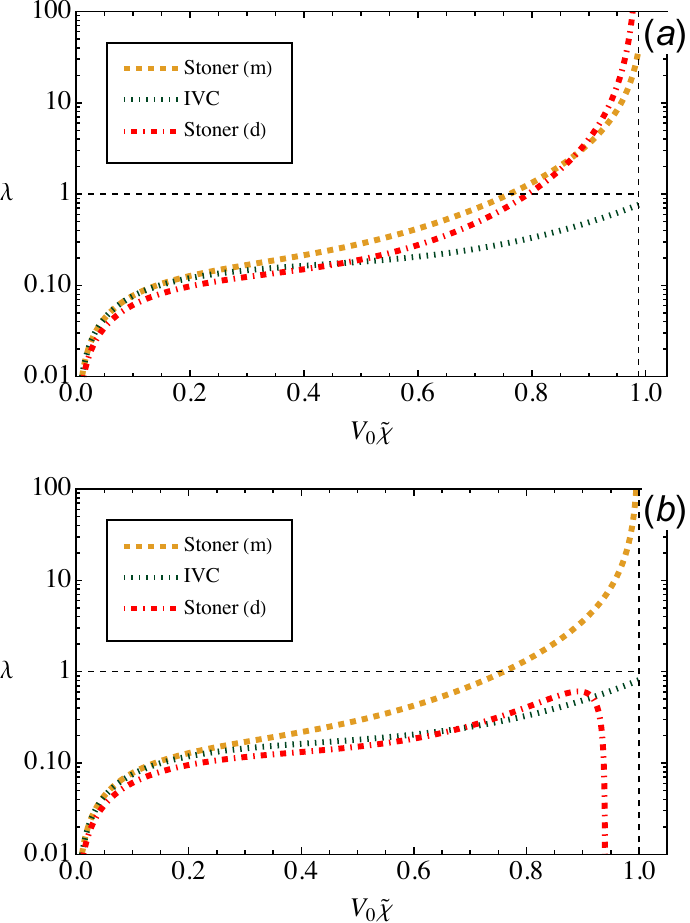}
\caption{Dependence of the eigenvalues for the correlated phases on $V_0 \tilde{\chi}$ for $SU(2)\times SU(2)$-symmetric interaction with $V_1=0.9V_0$ and (a) $N=10$, (b) $N=12$. The horizontal dashed line corresponds to $\lambda = 1$. The parquet approximation is applicable to the left of the vertical dashed line.
}
\label{fig:phase_diagram-lambda-parquet-su2-2}
\end{figure}

As another example, for $V_1=0.9 V_0$, we have no instabilities in the PP channel. However, there is an interplay between different Stoner instabilities.
For $N<9$, there is only Stoner instability in the density channel, while for $9 \le N \le 11$ there are regions with Stoner instability in the density and the magnetic (yellow dashed line) channels (the Stoner instability in the density channel emerges for lower temperatures), and for $N> 11$ there is only Stoner instability in the magnetic channel, since the eigenvalue $\lambda^{\text{Stoner}\,(d)}$ changes its asymptotic behavior.
The dependence of eigenvalues on $V_0 \tilde{\chi}(\beta,\mu)$ for $N=10,12$ is explicitly shown in Fig.~\ref{fig:phase_diagram-lambda-parquet-su2-2} for $V_1=0.9 V_0$.

In general, we can see that for 
\begin{equation*}
N \le \begin{cases}
\frac{3V_1-2V_0}{V_1-V_0},\,\,\,\,&V_1 >V_0, \\
\\
\frac{2V_0-V_1}{V_0-V_1},\,\,\,\,&V_1 <V_0 ,
\end{cases}
\end{equation*}
the eigenvalue $\lambda^{\text{Stoner}\,(d)} \rightarrow +\infty$ in the limit $\tilde{\chi} \rightarrow \tilde{\chi}^{(l)}$, while for 
\begin{equation*}
N > \begin{cases}
\frac{3V_1-2V_0}{V_1-V_0},\,\,\,\,&V_1 >V_0, \\
\\
\frac{2V_0-V_1}{V_0-V_1},\,\,\,\,&V_1 <V_0 ,
\end{cases}
\end{equation*}
the eigenvalue $\lambda^{\text{Stoner}\,(d)} \rightarrow -\infty$ in the limit $\tilde{\chi} \rightarrow \tilde{\chi}^{(l)}$. Here, the constant $\tilde{\chi}^{(l)}$ is defined in Eq.~(\ref{eq:chi-l-def}). Therefore, the Stoner instability in the density channel does not dominate at lower temperatures for the higher number of layers if the symmetry of $SU(4)$-symmetric interaction is broken. For $SU(4)$-symmetric interaction, it dominates for all $N \ge 3$.

\section{Conclusions}
\label{sec:summary}

We study the emergence of correlated phases in rhombohedral \(N\)-layer graphene using a two-valley Coulomb interaction model. We analytically derived Lindhard susceptibilities for both the intra- and intervalley channels based on an effective two-orbital low-energy \(k \cdot p\) model. In the next step, we estimated how the critical temperatures for the corresponding phase transitions depend on the number of layers, utilizing both the random-phase approximation and the parquet approximation.

Our predictions indicate that only the Stoner instability and the IVC phase may emerge within the RPA framework. Notably, instabilities in the PP channel cannot be captured by this approximation. We obtained a general scaling relationship for the critical temperature with respect to the number of layers, demonstrating the existence of an upper limit as \(N \rightarrow \infty\). Moreover, we predict that as the chemical potential increases from zero, the critical temperature decreases compared to the $\mu=0$ case. The susceptibilities exhibit non-monotonic behavior with the temperature.
Therefore, the phase transition from the normal state may occur while temperature increases from $0$, as illustrated in Figs.~\ref{fig:phase_diagram-rpa} and \ref{fig:phase_diagram-parquet}. 

We found that the parquet approximation identifies more correlated phases compared to the random phase approximation in our model description of rhombohedral \(N\)-layer graphene systems. We determined the dependence of the critical temperature on the number of layers, as shown in Eq.~(\ref{eq:temperature-scaling-su4}), and also established the upper limit for the critical temperature using the PA. Additionally, we analyzed the influence of the relationship between the interaction parameters \(V_{0}\) and \(V_{1}\) on the emergence of correlated phases. We predict that for \(SU(4)\)-symmetric interactions, the intervalley Stoner instability in the density channel always dominates. In contrast, we found that \(SU(2) \times SU(2)\)-symmetric interactions enable a wider range of instabilities to emerge, as shown in Fig.~\ref{fig:v1-crit}.

We also demonstrate that in the PH channel, there is a specific number of layers at which the eigenvalue for the Stoner instability in the density channel changes its asymptotic behavior. Consequently, other instabilities, such as the IVC phase or Stoner instability in the magnetic channel, may emerge for a larger number of layers. Furthermore, we found that the PDW phase may arise in the PP channel if \(V_{1} > V_{0}\). However, given that it is more likely for the intervalley repulsive interaction parameter to be smaller than the intravalley parameter, the existence of such a phase in real systems could be questionable.

\section*{Data availability}

The data and Wolfram Mathematica code used in this paper can be obtained from Ref.~\cite{data-correlated-phases}.

\begin{acknowledgments}

We would like to thank  V.\,P.~Gusynin  for numerous stimulating and enlightening discussions.
This work was supported by the Swiss National Science Foundation through the Ukrainian-Swiss Joint research project ``Transport and thermodynamic phenomena in low-dimensional materials with flat bands'' (grant No. IZURZ2\_224624).

\end{acknowledgments}

\appendix

\begin{widetext}

\section{Calculation of the bare susceptibilities}
\label{app:Kubo}

The retarded $(+)$ and advanced $(-)$ Green's functions read\\
\begin{equation}
G\left( \omega \pm i0,\mathbf{k}\right)=\frac{1}{\omega+\mu \pm i0 -H(\mathbf{k})}, 
\end{equation}
where $\omega$ is frequency, $\mathbf{k}$ is momentum, and $H(\mathbf{k})$ is a Hamiltonian.

The spectral function is defined as the difference between the advanced and retarded Green's functions at vanishing chemical potential
\begin{equation}
A\left( \omega ;\mathbf{k}\right)=\frac{i}{2\pi}\left[G\left( \omega + i0;\mathbf{k}\right)-G\left( \omega - i0;\mathbf{k}\right)\right]_{\mu=0}. 
\end{equation}

For the effective low-energy model, see Eq.~(\ref{eq:eff-hamiltonian}), the spectral function is given by 
\begin{equation}
A_{\tau}\left(\omega;\mathbf{k}\right)=\frac{1}{2}\left[\delta\left( \omega-\varepsilon_{\mathbf{k}} \right) + \delta\left( \omega+\varepsilon_{\mathbf{k}} \right) \right]+\frac{H_{\text{eff}}(\mathbf{k};\tau)}{2 \varepsilon_{\mathbf{k}} }\left[\delta\left( \omega-\varepsilon_{\mathbf{k}} \right) - \delta\left( \omega+\varepsilon_{\mathbf{k}} \right) \right].
\end{equation}
Here, $\varepsilon_{\mathbf{k}}$ is the dispersion relation of quasiparticles.

Equivalently, the Green's function $G_{\tau}\left(\omega;\mathbf{k}\right)$ can be represented in the following form:
\begin{equation}
G_{\tau}\left(\omega;\mathbf{k}\right)=\int_{-\infty}^{+\infty} d z \frac{A_{\tau}\left(z;\mathbf{k}\right)}{z-\omega-\mu} . \\
\end{equation}

Using 
\begin{equation*}
\frac{1}{\mathcal{N} }\sum_k\rightarrow \int \frac{d^2k}{\left(2\pi\right)^2},
\end{equation*}
Eqs.~(\ref{eq:PH-0-def}) and (\ref{eq:PP-0-def}) can be rewritten as follows:

\begin{equation}
\chi^{(0),\text{PH}}_{\tau_1,\tau_2}=- T  \sum_{n=-\infty}^{+\infty}\int \frac{d^2k}{\left(2\pi\right)^2} \int_{-\infty}^{+\infty} d \omega \int_{-\infty}^{+\infty} d \omega'  \frac{1}{i\omega_n +\mu - \omega  }\frac{1}{i\omega_n +\mu- \omega'  } \text{tr} \left[ A_{\tau_1}\left(\omega;\mathbf{k}\right)  A_{\tau_2}\left(\omega';\mathbf{k}\right) \right] ,\\
\end{equation}
\begin{equation}
\chi^{(0),\text{PP}}_{\tau_1,\tau_2}=- T  \sum_{n=-\infty}^{+\infty}\int \frac{d^2k}{\left(2\pi\right)^2} \int_{-\infty}^{+\infty} d \omega \int_{-\infty}^{+\infty} d \omega'  \frac{1}{i\omega_n +\mu- \omega  }\frac{1}{-i\omega_n + \mu - \omega'} \text{tr} \left[ A_{\tau_1}\left(\omega;\mathbf{k}\right)  A_{\tau_2}\left(\omega';-\mathbf{k}\right) \right] .\\
\end{equation}
After performing summation over the fermionic Matsubara frequencies $\omega_{n}= \left(2 n+1 \right) \pi T$,
we finally obtained
\begin{equation}
\label{eq:app-susc-ph-def-1}
\chi^{(0),\text{PH}}_{\tau_1,\tau_2}= \int_{-\infty}^{+\infty} d \omega \int_{-\infty}^{+\infty} d \omega' \frac{f\left( \omega \right)- f\left( \omega' \right)}{\omega'-\omega} \int \frac{d^2k}{\left(2\pi\right)^2} \text{tr} \left[ A_{\tau_1}\left(\omega;\mathbf{k}\right)  A_{\tau_2}\left(\omega';\mathbf{k}\right) \right] ,\\
\end{equation}
\begin{equation}
\label{eq:app-susc-pp-def-1}
\chi^{(0),\text{PP}}_{\tau_1,\tau_2}=-   \int_{-\infty}^{+\infty} d \omega \int_{-\infty}^{+\infty} d \omega' \frac{f\left( \omega \right)- f\left( \omega' \right)}{\omega'-\omega} \int \frac{d^2k}{\left(2\pi\right)^2} \text{tr} \left[ A_{\tau_1}\left(\omega;\mathbf{k}\right)  A_{\tau_2}\left(-\omega';-\mathbf{k}\right) \right] .\\
\end{equation}
where $f(\omega)=1/\left[ e^{(\omega-\mu)/T}+1 \right]$ is the Fermi-Dirac distribution function. 
For generality, we include the term $-U\sigma_{z}/2$ that corresponds to interlayer asymmetry between the upper and lower layers to the effective model Hamiltonian, presented in Eq.~(\ref{eq:eff-hamiltonian}). The traces in Eqs.~(\ref{eq:app-susc-ph-def-1}) and (\ref{eq:app-susc-pp-def-1}) are given by
\begin{equation}
\begin{aligned}
\text{tr} \left[   A_{\tau_1}(\omega;\mathbf{k}) A_{\tau_2}(\omega';\mathbf{k}) \right] &= \frac{1}{2} \left[ \delta\left( \omega - \varepsilon_{\mathbf{k}} \right) + \delta\left( \omega + \varepsilon_{\mathbf{k}} \right)\right]  \left[ \delta\left( \omega' - \varepsilon_{\mathbf{k} } \right) + \delta\left( \omega' + \varepsilon_{\mathbf{k} } \right)\right]  \\
&+\frac{1}{2 \varepsilon_{\mathbf{k}}^2 }  \left[ \delta\left( \omega - \varepsilon_{\mathbf{k}} \right) - \delta\left( \omega + \varepsilon_{\mathbf{k}} \right)\right]  \left[ \delta\left( \omega' - \varepsilon_{\mathbf{k} } \right) - \delta\left( \omega' + \varepsilon_{\mathbf{k} } \right)\right]   \\
&\times \left[  \left(\tau_1 \tau_2 \sin ^2(N \phi )+\cos ^2(N \phi )\right) \left( \varepsilon_{\mathbf{k}}^2 -\frac{U^2}{4} \right)  +\frac{U^2}{4}\right],  \\
\end{aligned}
\end{equation}
\begin{equation}
\begin{aligned}
\text{tr} \left[   A_{\tau_1}(\omega;\mathbf{k}) A_{\tau_2}(-\omega';-\mathbf{k}) \right] &= \frac{1}{2} \left[ \delta\left( \omega - \varepsilon_{\mathbf{k}} \right) + \delta\left( \omega + \varepsilon_{\mathbf{k}} \right)\right]  \left[ \delta\left( \omega' - \varepsilon_{\mathbf{k}} \right) + \delta\left( \omega' + \varepsilon_{\mathbf{k} } \right)\right]  \\
&- \frac{1}{2 \varepsilon_{\mathbf{k}}^2 }  \left[ \delta\left( \omega - \varepsilon_{\mathbf{k}} \right) - \delta\left( \omega + \varepsilon_{\mathbf{k}} \right)\right]  \left[ \delta\left( \omega' - \varepsilon_{\mathbf{k}} \right) - \delta\left( \omega' + \varepsilon_{\mathbf{k} } \right)\right]   \\
&\times \left[ (-1)^N  \left(\tau_1 \tau_2 \sin ^2(N \phi )+\cos ^2(N \phi )\right)  \left( \varepsilon_{\mathbf{k}}^2 -\frac{U^2}{4} \right) +\frac{U^2}{4}\right]  \\
\end{aligned}
\end{equation}
with modified dispersion $\varepsilon_{\mathbf{k}}= \sqrt{g_{N}^2 k^{2N}  + U^2/ 4} $.

Therefore, for intravalley particle-hole and particle-particle susceptibilities ($\tau_1=\tau_2$), we obtain
\begin{equation}
\label{eq:intra-ph-susceptibility}
\begin{aligned}
\chi_{\tau,\tau}^{(0),\text{PH}} &=\frac{1}{4\pi N \left(g_N\right)^{\frac{2}{N}}}  \int_{\frac{|U|}{2}}^{+\infty}d\varepsilon\,\frac{\varepsilon}{\left(\varepsilon^2 -\frac{U^2}{4} \right)^{1-\frac{1}{N} }} \,  \sum_{\lambda} \frac{2 \beta e^{\beta \left( \varepsilon -\lambda \mu \right)}}{\left[ e^{\beta \left( \varepsilon -\lambda \mu \right)}+ 1 \right]^2}  ,\\
\end{aligned}
\end{equation}
\begin{equation}
\label{eq:intra-pp-susceptibility}
\begin{aligned}
\chi^{(0),\text{PP}}_{\tau,\tau}&=-\frac{1 }{4\pi N \left(g_N\right)^{\frac{2}{N}}}  \int_{\frac{|U|}{2}}^{+\infty}d\varepsilon\,\frac{\varepsilon}{\left(\varepsilon^2 -\frac{U^2}{4} \right)^{1-\frac{1}{N} }} \, \Bigg\{ \sum_{\lambda} \frac{ \beta e^{\beta \left( \varepsilon -\lambda \mu \right)}}{\left[ e^{\beta \left( \varepsilon -\lambda \mu \right)}+ 1 \right]^2}  \left( 1 - (-1)^N \right)  \left(1 -\frac{U^2}{4\varepsilon^2 } \right) \\
&+\frac{f\left( -\varepsilon \right)-f\left( \varepsilon \right)}{\varepsilon } \left[ \left( 1 + (-1)^N \right) + \left( 1 - (-1)^N \right) \frac{U^2}{4\varepsilon^2 }\right]  \Bigg\} .\\
\end{aligned}
\end{equation}
For intervalley susceptibilities ($\tau_1=-\tau_2$), we obtain
\begin{equation}
\label{eq:inter-ph-susceptibility}
\chi_{\tau, -\tau}^{(0),\text{PH}} =\frac{1}{4\pi N \left(g_N\right)^{\frac{2}{N}}}  \int_{\frac{|U|}{2}}^{+\infty}d\varepsilon\,\frac{\varepsilon}{\left(\varepsilon^2 -\frac{U^2}{4} \right)^{1-\frac{1}{N} }} \, \Bigg\{ \sum_{\lambda} \frac{\beta e^{\beta \left( \varepsilon -\lambda \mu \right)}}{\left[ e^{\beta \left( \varepsilon -\lambda \mu \right)}+ 1 \right]^2}  \left[ 1+\frac{U^2}{4\varepsilon^2 }\right] +\frac{f\left( -\varepsilon \right)-f\left( \varepsilon \right)}{\varepsilon } \left[ 1-\frac{U^2}{4\varepsilon ^2 }\right]  \Bigg\} ,\\
\end{equation}
\begin{equation}
\label{eq:inter-pp-susceptibility}
\chi^{(0),\text{PP}}_{\tau ,-\tau }=-\frac{1 }{4\pi N \left(g_N\right)^{\frac{2}{N}}}  \int_{\frac{|U|}{2}}^{+\infty}d\varepsilon\,\frac{\varepsilon}{\left(\varepsilon^2 -\frac{U^2}{4} \right)^{1-\frac{1}{N} }} \, \Bigg\{ \sum_{\lambda} \frac{\beta e^{\beta \left( \varepsilon -\lambda \mu \right)}}{\left[ e^{\beta \left( \varepsilon -\lambda \mu \right)}+ 1 \right]^2}  \left[ 1 - \frac{U^2}{4\varepsilon^2 }\right] +\frac{f\left( -\varepsilon \right)-f\left( \varepsilon \right)}{\varepsilon } \left[ 1 + \frac{U^2}{4\varepsilon^2 }\right]  \Bigg\} .\\
\end{equation}
The numerical integration results at $U \neq 0$ and $N=3$ for the particle-hole susceptibility, given by Eqs.~(\ref{eq:intra-ph-susceptibility}) and (\ref{eq:inter-ph-susceptibility}), and the particle-particle susceptibility, given by Eqs.~(\ref{eq:intra-pp-susceptibility}) and (\ref{eq:inter-pp-susceptibility}), are shown in Figs.~\ref{fig:irr-ph-susceptibility} and \ref{fig:irr-pp-susceptibility} respectively. 
\begin{figure}[h]
\includegraphics[width=.9\textwidth]{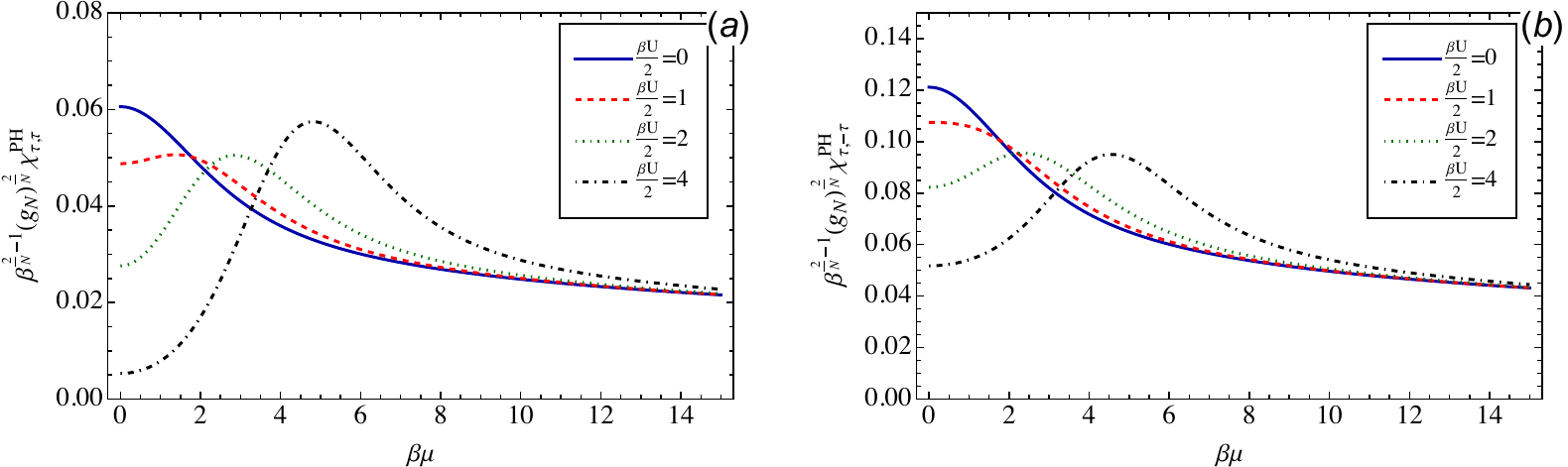}
\caption{Dependence of the particle-hole susceptibilities, given by Eqs.~(\ref{eq:intra-ph-susceptibility}) and (\ref{eq:inter-ph-susceptibility}), for $N=3$ on $\beta \mu$ for $U=0$ and $U\neq0$: (a) intravalley case and (b) intervalley case.}
\label{fig:irr-ph-susceptibility}
\end{figure} 

\begin{figure}[h]
\includegraphics[width=.9\textwidth]{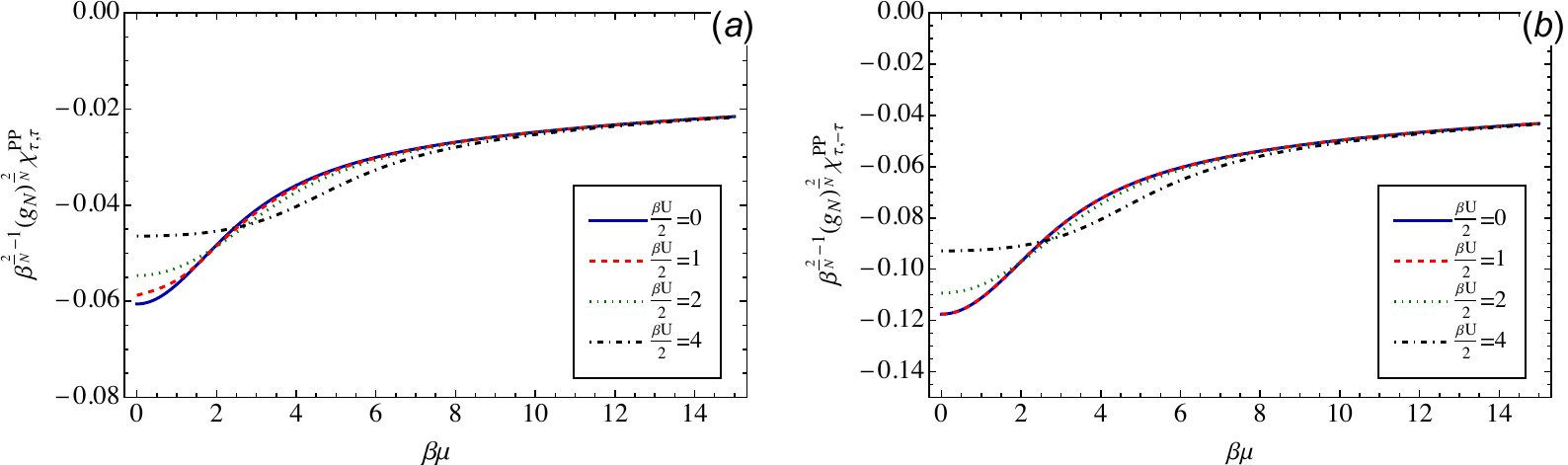}
\caption{Dependence of the particle-particle susceptibilities, given by Eqs.~(\ref{eq:intra-pp-susceptibility}) and (\ref{eq:inter-pp-susceptibility}), for $N=3$ on $\beta \mu$ for $U=0$ and $U\neq0$: (a) intravalley case and (b) intervalley case.}
\label{fig:irr-pp-susceptibility}
\end{figure}

Non-zero displacement field $U$ makes the dependence of the intra- and intervalley particle-hole susceptibilities on $\beta \mu$ non-monotonic. For the particle-particle susceptibilities, nontrivial $U$ increases the absolute minimum at $\beta \mu =0$, but the monotonic behavior on $\beta \mu$ is preserved. We also present the dependence of $\tilde{\chi}$ on temperature $T$ for several values of chemical potential and $N=3$ in Fig.~\ref{fig:tilde-chi-t} (a). Table~\ref{tab:table-num} is used for numerical computations of Eq.~(\ref{eq:tilde-chi-def}).
\begin{figure}[h]
\includegraphics[width=.9\textwidth]{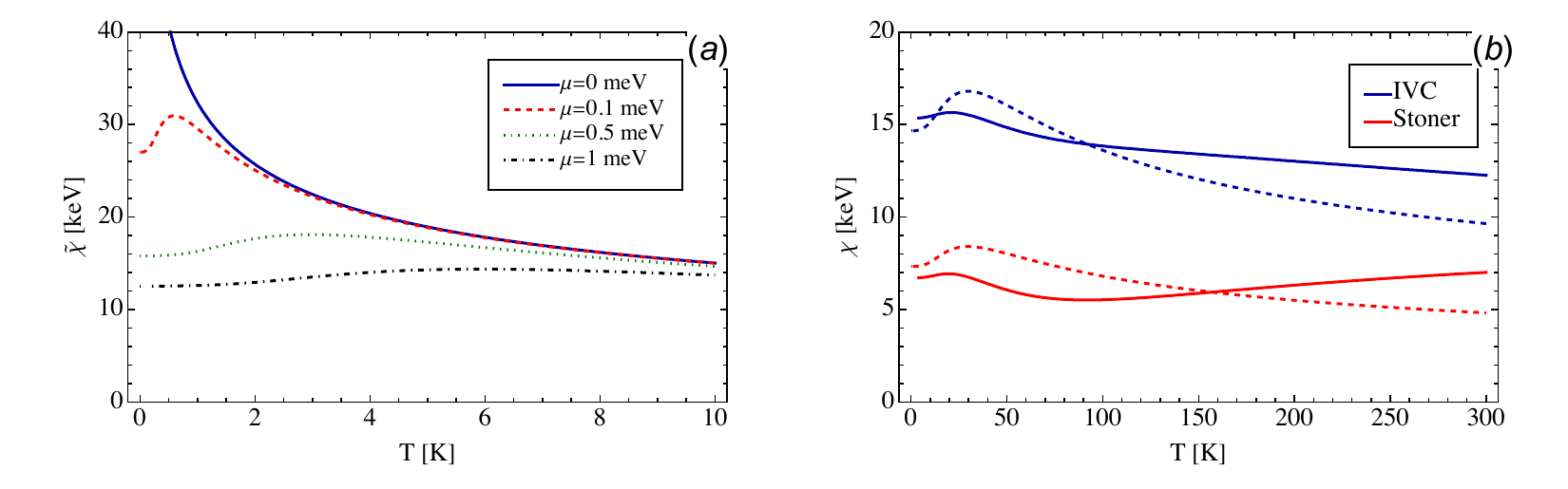}
\caption{(a) Dependence of $\tilde{\chi}$ on $T$ for several values of $\mu$ and $N=3$. (b) The dependence of intra- and intervalley susceptibilities for the full model of rhombohedral trilayer graphene on temperature $T$ at $\mu= 5 ~\mbox{meV}$. Blue and red dashed lines correspond to inter- and intravalley susceptibilities at $\mu= 5 ~\mbox{meV}$ and $U=0$, given by Eqs.~(\ref{eq:inter-ph-susceptibility}) and (\ref{eq:intra-ph-susceptibility}), respectively.}
\label{fig:tilde-chi-t}
\end{figure}

In Fig.~\ref{fig:tilde-chi-t} (b), we present the results of calculations for intra- and intervalley susceptibilities for the full model of rhombohedral trilayer graphene based on a realistic Hamiltonian parametrization~\cite{PhysRevB.105.115126}
\begin{equation}
\label{eq:RTG-DFT-model}
\centering
    H_{\text{RTG}}(\mathbf{k})\
=\begin{pmatrix}
  u_d & \gamma_0 f&\gamma_4 f^{*}&\gamma_1&0&0 \\ 
  \gamma_0 f^{*} & \eta+u_d&\gamma_3 f&\gamma_4 f^{*}&\gamma_6&0\\
  \gamma_4 f&\gamma_3 f^{*}&u_m&\gamma_0 f&\gamma_4 f^{*}&\gamma_1\\
  \gamma_1&\gamma_4 f&\gamma_0 f^{*}&u_m&\gamma_3 f&\gamma_4 f^{*}\\
  0&\gamma_6&\gamma_4 f&\gamma_3 f^{*}&\eta-u_d&\gamma_0 f\\
  0&0&\gamma_1&\gamma_4 f&\gamma_0 f^{*}&-u_d
\end{pmatrix} ,
\end{equation}
where we consider single-particle Bloch states with momenta $\mathbf{k} = (k_x, k_y)$, measured from the $K$ and $K'$ valleys, constructed from the carbon $p_z$ atomic orbitals located at the rhombohedral trilayer graphene sites ($A_1, B_1, A_2, B_2, A_3, B_3$). The linearized nearest-neighbor structure factor is given by $f = -\left(\sqrt{3}a/2\right)\left(\tau k_x - i k_y\right)$.
The parameters ${\gamma_i}$ represent orbital hopping amplitudes between sites in the SWM notations. Electrostatic potentials from an applied displacement field are included via the on-site energies $u_d$, $u_m$, and $-u_d$, with $2u_d$ corresponding to the potential energy difference between the outermost layers. Additionally, an on-site potential $\eta$ is applied to sites that participate in vertical hybridization mediated by the $\gamma_6$ hopping term.

The position of a maximum of $\tilde{\chi}$, presented in Fig.~\ref{fig:tilde-chi-t} (a), is at $T=t\mu$, where $t$ is the solution of Eq.~(\ref{eq:sol1-axillary-1}). The dependence of $t$ on $N$ is presented in Fig.~\ref{fig:t-n-axillary}.
\begin{figure}[h]
\includegraphics[width=.45\textwidth]{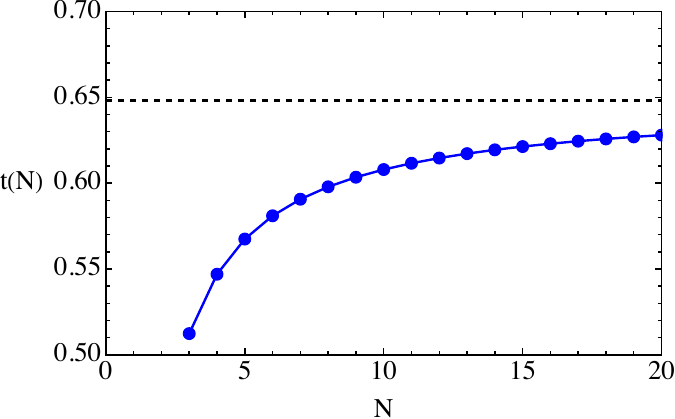}
\caption{Dependence of temperature $T$ in units of the chemical potential, at which $\tilde{\chi}$ reaches its maximum, on number of layers $N$. The black dashed line corresponds to the solution of Eq.~(\ref{eq:sol1-axillary-1}).}
\label{fig:t-n-axillary}
\end{figure}

\section{Explicit expressions for the channel irreducible vertices in the parquet approximation}
\label{app:explicit-vertices}

To obtain the equation for the emerging correlated phase, we need to separate the channel-irreducible vertices $\Gamma$ at the level of Eq.~(\ref{eq:app_ir_vertex}) in the singlet, triplet, magnetic, and density channels using the definition Eq.~(\ref{eq:spin-channel-sep}). In the singlet channel, we obtain the following equation for channel-irreducible vertex:
\begin{equation}
\label{eq:gamma-s-irr}
\begin{aligned}
&\Gamma^{s}\left( k , k^{\prime},q\right)=\Gamma^{\text{PP}}_{\uparrow \uparrow \downarrow \downarrow}\left( k , k^{\prime},q\right)-\Gamma^{\text{PP}}_{\uparrow  \downarrow \downarrow \uparrow }\left( k , k^{\prime},q\right)=\Lambda^{\text{PP}}_{\uparrow \uparrow \downarrow \downarrow} -\Lambda^{\text{PP}}_{\uparrow \downarrow \downarrow \uparrow  } \\ - &\left[ \sum_{ \left\{ \sigma_i \right\} } \Gamma^{\text{PH}}_{\uparrow \uparrow \sigma_1 \sigma_2}  \chi^{(0),\text{PH}}_{\sigma_2 \sigma_1 \sigma_4 \sigma_3  }F^{\text{PH}}_{\sigma_3 \sigma_4 \downarrow \downarrow}\right] \left( k , k^{\prime},q-k - k^{\prime}\right)-  \left[ \sum_{ \left\{ \sigma_i \right\} }  \Gamma^{\text{vPH}}_{\uparrow  \sigma_1  \sigma_2 \downarrow}  \chi^{(0),\text{vPH}}_{\sigma_1 \sigma_3  \sigma_4 \sigma_2 }F^{\text{vPH}}_{\sigma_3 \uparrow \downarrow \sigma_4  } \right] \left( k , q-k^{\prime},-k + k^{\prime}\right)\\+& \left[ \sum_{ \left\{ \sigma_i \right\} } \Gamma^{\text{PH}}_{\uparrow \downarrow \sigma_1 \sigma_2}  \chi^{(0),\text{PH}}_{\sigma_2 \sigma_1 \sigma_4 \sigma_3  }F^{\text{PH}}_{\sigma_3 \sigma_4 \uparrow \downarrow} \right] \left( k , k^{\prime},q-k - k^{\prime}\right) + \left[ \sum_{\left\{ \sigma_i \right\}}  \Gamma^{\text{vPH}}_{\uparrow  \sigma_1 \sigma_2 \uparrow }  \chi^{(0),\text{vPH}}_{\sigma_1 \sigma_3 \sigma_4 \sigma_2  }F^{\text{vPH}}_{\sigma_3 \downarrow \downarrow \sigma_4  }\right]\left( k , q-k^{\prime},-k + k^{\prime}\right) 
\\=&\Lambda^{s} - \left[ \frac{1}{2} \Gamma^{d}  \chi^{(0),\text{PH}} F^{d}- \frac{3}{2} \Gamma^{m}  \chi^{(0),\text{PH}}F^{m}\right]  \left( k , k^{\prime},q-k - k^{\prime}\right) - \left[ \frac{1}{2} \Gamma^{d}  \chi^{(0),\text{PH}} F^{d}- \frac{3}{2} \Gamma^{m}  \chi^{(0),\text{PH}}F^{m}\right]\left( k , q-k^{\prime},-k + k^{\prime}\right) . \\
\end{aligned}
\end{equation}
Since the fully reducible vertices in the magnetic and density channels are given by
\begin{equation}
\label{eq:app-red-full-m-d}
\begin{aligned}
\Phi^{d}&=-\Gamma^{d}\chi^{d,(0),\text{PH}}F^{d} ,\\
\Phi^{m}&=-\Gamma^{m}\chi^{m,(0),\text{PH}}F^{m} ,\\
\end{aligned}
\end{equation}
then after substituting Eq.~(\ref{eq:app-red-full-m-d}) in Eq.~(\ref{eq:gamma-s-irr}), we arrive at Eq.~(\ref{eq:irr-vertex-s}). 

Similarly, in the triplet channel, we obtain the following equation for channel-irreducible vertex:
\begin{equation}
\begin{aligned}
&\Gamma^{t}\left( k , k^{\prime},q\right)=\Gamma^{\text{PP}}_{\uparrow \uparrow \downarrow \downarrow}\left( k , k^{\prime},q\right)+\Gamma^{\text{PP}}_{\uparrow  \downarrow \downarrow \uparrow }\left( k , k^{\prime},q\right)=\Lambda^{\text{PP}}_{\uparrow \uparrow \downarrow \downarrow} +\Lambda^{\text{PP}}_{\uparrow \downarrow \downarrow \uparrow  } \\ -&\left[ \sum_{ \left\{ \sigma_i \right\} } \Gamma^{\text{PH}}_{\uparrow \uparrow \sigma_1 \sigma_2}  \chi^{(0),\text{PH}}_{\sigma_2 \sigma_1 \sigma_4 \sigma_3  }F^{\text{PH}}_{\sigma_3 \sigma_4 \downarrow \downarrow}\right] \left( k , k^{\prime},q-k - k^{\prime}\right)-  \left[ \sum_{ \left\{ \sigma_i \right\} }  \Gamma^{\text{vPH}}_{\uparrow  \sigma_1  \sigma_2 \downarrow}  \chi^{(0),\text{vPH}}_{\sigma_1 \sigma_3  \sigma_4 \sigma_2 }F^{\text{vPH}}_{\sigma_3 \uparrow \downarrow \sigma_4  } \right] \left( k , q-k^{\prime},-k + k^{\prime}\right)\\-& \left[ \sum_{ \left\{ \sigma_i \right\} } \Gamma^{\text{PH}}_{\uparrow \downarrow \sigma_1 \sigma_2}  \chi^{(0),\text{PH}}_{\sigma_2 \sigma_1 \sigma_4 \sigma_3  }F^{\text{PH}}_{\sigma_3 \sigma_4 \uparrow \downarrow} \right] \left( k , k^{\prime},q-k - k^{\prime}\right) - \left[ \sum_{\left\{ \sigma_i \right\}}  \Gamma^{\text{vPH}}_{\uparrow  \sigma_1 \sigma_2 \uparrow }  \chi^{(0),\text{vPH}}_{\sigma_1 \sigma_3 \sigma_4 \sigma_2  }F^{\text{vPH}}_{\sigma_3 \downarrow \downarrow \sigma_4  }\right]\left( k , q-k^{\prime},-k + k^{\prime}\right) 
\\=&\Lambda^{t} - \left[ \frac{1}{2} \Gamma^{d}  \chi^{(0),\text{PH}} F^{d}+ \frac{1}{2} \Gamma^{m}  \chi^{(0),\text{PH}}F^{m}\right]  \left( k , k^{\prime},q-k - k^{\prime}\right) + \left[ \frac{1}{2} \Gamma^{d}  \chi^{(0),\text{PH}} F^{d}+ \frac{1}{2} \Gamma^{m}  \chi^{(0),\text{PH}}F^{m}\right]\left( k , q-k^{\prime},-k + k^{\prime}\right) . \\
\end{aligned}
\end{equation}
After substituting the expressions for $\Phi^{m}$ and $\Phi^{d}$ from Eq.~(\ref{eq:app-red-full-m-d}), we arrive at Eq.~(\ref{eq:irr-vertex-t}).

In the parquet approximation, we obtain the following approximated expressions for the singlet and triplet channel-irreducible vertices using the expressions for fully reducible vertices $\Phi$, given by Eq.~(\ref{eq:app_reducible_vertex-approx}):

\begin{equation}
\label{eq:app-irr-vertex-s-flex}
\begin{aligned}
\Gamma^{s}&= 
\left[2V_0+ \frac{3 V_0^2 \chi_{\tau, \tau}^{(0),\text{PH}}}{1-V_0 \chi_{\tau, \tau}^{(0),\text{PH}}}-\chi_{\tau, \tau}^{(0),\text{PH}} \frac{\left(V_0^2 -4 V_1^2\right)V_0 \chi_{\tau, \tau}^{(0),\text{PH}} +V_0^2 +4 V_1^2}{(V_0 \chi_{\tau, \tau}^{(0),\text{PH}} +1 )^2-4 V_1^2 (\chi_{\tau, \tau}^{(0),\text{PH}})^2}\right]
\left(
\begin{array}{cccc}
1& 0 & 0 & 0 \\
 0 & 0 & 0 & 0 \\
 0 & 0& 0& 0 \\
0 & 0 & 0 & 1\\
\end{array}
\right) \\
&+V_1 \left[1+\frac{V_1 \chi_{\tau, -\tau}^{(0),\text{PH}} }{1-V_1\chi_{\tau, -\tau}^{(0),\text{PH}} }-\frac{\chi_{\tau, \tau}^{(0),\text{PH}} \left(2 V_0 + V_0^2 \chi_{\tau, \tau}^{(0),\text{PH}} -4 V_1^2 \chi_{\tau, \tau}^{(0),\text{PH}} \right)}{(V_0 \chi_{\tau, \tau}^{(0),\text{PH}} +1 )^2-4 V_1^2 (\chi_{\tau, \tau}^{(0),\text{PH}})^2}\right]
\left(
\begin{array}{cccc}
 0 & 0 & 0 & 0 \\
 0 &1& 1 & 0 \\
 0 &1& 1 & 0 \\
0& 0 & 0 & 0 \\
\end{array}
\right),
\end{aligned}
\end{equation} 
\begin{equation}
\label{eq:app-irr-vertex-t-flex}
\Gamma^{t}=
V_1 \left[1+\frac{V_1 \chi_{\tau, -\tau}^{(0),\text{PH}} }{1-V_1\chi_{\tau, -\tau}^{(0),\text{PH}} }-\frac{\chi_{\tau, \tau}^{(0),\text{PH}} \left(2 V_0 + V_0^2 \chi_{\tau, \tau}^{(0),\text{PH}} -4 V_1^2 \chi_{\tau, \tau}^{(0),\text{PH}} \right)}{(V_0 \chi_{\tau, \tau}^{(0),\text{PH}} +1 )^2-4 V_1^2 (\chi_{\tau, \tau}^{(0),\text{PH}})^2}\right]
\left(
\begin{array}{cccc}
 0 & 0 & 0 & 0 \\
 0 & 1 & -1& 0 \\
 0 & -1 & 1 & 0 \\
0& 0 & 0 & 0 \\
\end{array}
\right).
\end{equation}

From Eq.~(\ref{eq:app_ir_vertex}), we obtain the following expressions for the channel-irreducible vertex in the magnetic channel:
\begin{equation}
\begin{aligned}
&\Gamma^{m}\left( k , k^{\prime},q\right)=\Gamma^{\text{PH}}_{\uparrow\uparrow \downarrow  \downarrow}\left( k , k^{\prime},q\right) -\Gamma^{\text{PH}}_{\uparrow \uparrow \uparrow \uparrow}\left( k , k^{\prime},q\right) = \Lambda^{\text{PH}}_{\uparrow  \uparrow  \downarrow \downarrow} - \Lambda^{\text{PH}}_{\uparrow \uparrow \uparrow \uparrow}\\
-& \frac{1}{2} \left[ \sum_{\left\{ \sigma_{i} \right\} }\Gamma^{\text{PP}}_{\uparrow  \sigma_1 \downarrow \sigma_2 }  \chi^{(0),\text{PP}}_{\sigma_2 \sigma_4 \sigma_1 \sigma_3  }F^{\text{PP}}_{\sigma_3 \uparrow \sigma_4 \downarrow  }\right] \left( k , k^{\prime},q+ k + k^{\prime}\right) + \left[ \sum_{\left\{ \sigma_{i} \right\} }  \Gamma^{\text{vPH}}_{\uparrow  \sigma_1 \sigma_2 \downarrow }  \chi^{(0),\text{vPH}}_{\sigma_1 \sigma_3 \sigma_4 \sigma_2  }F^{\text{vPH}}_{\sigma_3 \uparrow \downarrow \sigma_4  }\right] \left( k , q+k,- k + k^{\prime}\right)  \\
+&\frac{1}{2} \left[ \sum_{\left\{ \sigma_{i} \right\} } \Gamma^{\text{PP}}_{\uparrow  \sigma_1 \uparrow \sigma_2 }  \chi^{(0),\text{PP}}_{\sigma_2 \sigma_4 \sigma_1 \sigma_3  }F^{\text{PP}}_{\sigma_3 \uparrow  \sigma_4 \uparrow  }\right] \left( k , k^{\prime},q + k + k^{\prime}\right) - \left[ \sum_{\left\{ \sigma_{i} \right\}}  \Gamma^{\text{vPH}}_{\uparrow  \sigma_1 \sigma_2 \uparrow }  \chi^{(0),\text{vPH}}_{\sigma_1 \sigma_3 \sigma_4 \sigma_2  }F^{\text{vPH}}_{\sigma_3 \uparrow   \uparrow \sigma_4} \right] \left( k , q+k,- k + k^{\prime}\right) \\
=&\Lambda^{m} + \frac{1}{2} \left[ \frac{1}{2} \Gamma^{t}  \chi^{(0),\text{PP}} F^{t} - \frac{1}{2} \Gamma^{s}  \chi^{(0),\text{PP}} F^{s}  \right] \left( k , k^{\prime},q+ k + k^{\prime}\right) - \left[ \frac{1}{2} \Gamma^{d}  \chi^{(0),\text{PH}} F^{d} - \frac{1}{2} \Gamma^{m}  \chi^{(0),\text{PH}}F^{m}\right] \left( k , q+k,- k + k^{\prime}\right).,  \\
\end{aligned}
\end{equation}
Also from Eq.~(\ref{eq:app_ir_vertex}), the equation for the channel-irreducible vertex in the density channel reads:
\begin{equation}
\begin{aligned}
&\Gamma^{d}\left( k , k^{\prime},q\right)=\Gamma^{\text{PH}}_{\uparrow \uparrow \uparrow \uparrow}\left( k , k^{\prime},q\right) + \Gamma^{\text{PH}}_{\uparrow\uparrow \downarrow  \downarrow}\left( k , k^{\prime},q\right) =\Lambda^{\text{PH}}_{\uparrow \uparrow \uparrow \uparrow} + \Lambda^{\text{PH}}_{\uparrow  \uparrow  \downarrow \downarrow}\\
-&\frac{1}{2} \left[ \sum_{\left\{ \sigma_{i} \right\} } \Gamma^{\text{PP}}_{\uparrow  \sigma_1 \uparrow \sigma_2 }  \chi^{(0),\text{PP}}_{\sigma_2 \sigma_4 \sigma_1 \sigma_3  }F^{\text{PP}}_{\sigma_3 \uparrow  \sigma_4 \uparrow  }\right] \left( k , k^{\prime},q + k + k^{\prime}\right) - \left[ \sum_{\left\{ \sigma_{i} \right\}}  \Gamma^{\text{vPH}}_{\uparrow  \sigma_1 \sigma_2 \uparrow }  \chi^{(0),\text{vPH}}_{\sigma_1 \sigma_3 \sigma_4 \sigma_2  }F^{\text{vPH}}_{\sigma_3 \uparrow   \uparrow \sigma_4} \right] \left( k , q+k,- k + k^{\prime}\right) \\
-& \frac{1}{2} \left[ \sum_{\left\{ \sigma_{i} \right\} }\Gamma^{\text{PP}}_{\uparrow  \sigma_1 \downarrow \sigma_2 }  \chi^{(0),\text{PP}}_{\sigma_2 \sigma_4 \sigma_1 \sigma_3  }F^{\text{PP}}_{\sigma_3 \uparrow \sigma_4 \downarrow  }\right] \left( k , k^{\prime},q+ k + k^{\prime}\right) - \left[ \sum_{\left\{ \sigma_{i} \right\} }  \Gamma^{\text{vPH}}_{\uparrow  \sigma_1 \sigma_2 \downarrow }  \chi^{(0),\text{vPH}}_{\sigma_1 \sigma_3 \sigma_4 \sigma_2  }F^{\text{vPH}}_{\sigma_3 \uparrow \downarrow \sigma_4  }\right] \left( k , q+k,- k + k^{\prime}\right)  \\
=&\Lambda^{d} - \frac{1}{2} \left[ \frac{3}{2} \Gamma^{t}  \chi^{(0),\text{PP}} F^{t} + \frac{1}{2} \Gamma^{s}  \chi^{(0),\text{PP}} F^{s}  \right] \left( k , k^{\prime},q+ k + k^{\prime}\right) + \left[ \frac{1}{2} \Gamma^{d}  \chi^{(0),\text{PH}} F^{d}+ \frac{3}{2} \Gamma^{m}  \chi^{(0),\text{PH}}F^{m}\right] \left( k , q+k,- k + k^{\prime}\right)  . \\
\end{aligned}
\end{equation}
Since the fully reducible vertices in the singlet and triplet channels are given by
\begin{equation}
\label{eq:app-red-full-s-t}
\begin{aligned}
\Phi^{s}&=-\frac{1}{2}\Gamma^{s}\chi^{s,(0),\text{PP}}F^{s} ,\\
\Phi^{t}&=-\frac{1}{2}\Gamma^{t}\chi^{t,(0),\text{PP}}F^{t} ,\\
\end{aligned}
\end{equation}
then we arrive at Eqs.~(\ref{eq:irr-vertex-m}) and (\ref{eq:irr-vertex-d}).

Similarly, we obtain the following approximated expressions for the channel-irreducible vertices in magnetic and density channels using the expressions for fully reducible vertices $\Phi$, given by Eq.~(\ref{eq:app_reducible_vertex-approx}):
\begin{equation}
\label{eq:app-irr-vertex-m-flex}
\begin{aligned}
\Gamma^{m}&= \frac{1}{(V_0 \chi_{\tau, \tau}^{(0),\text{PH}} +1 )^2-4 V_1^2 (\chi_{\tau, \tau}^{(0),\text{PH}})^2} \left(
\begin{array}{cccc}
\frac{V_0+(V_0^2 -2V_1^2)  \chi_{\tau, \tau}^{(0),\text{PH}} }{1 - V_0  \chi_{\tau, \tau}^{(0),\text{PH}} }& 0 & 0 & 0 \\
 0 & V_1 & 0 & 0 \\
 0 & 0& V_1& 0 \\
0 & 0 & 0 & \frac{V_0+(V_0^2 -2V_1^2)  \chi_{\tau, \tau}^{(0),\text{PH}} }{1 - V_0  \chi_{\tau, \tau}^{(0),\text{PH}} }\\
\end{array}
\right)\\
&-\frac{V_0^2 \chi_{\tau,\tau}^{(0),\text{PP}}}{V_0 \chi_{\tau,\tau}^{(0),\text{PP}}-1} \left(
\begin{array}{cccc}
1& 0 & 0 & 0 \\
 0 & 0 & 0 & 0 \\
 0 & 0& 0& 0 \\
0 & 0 & 0 & 1\\
\end{array}
\right) -\frac{V_1^2 \chi_{\tau,-\tau}^{(0),\text{PP}}}{V_1 \chi_{\tau,-\tau}^{(0),\text{PP}}-1} \left(
\begin{array}{cccc}
0& 0 & 0 & 0 \\
 0 & 1 & 0 & 0 \\
 0 & 0& 1& 0 \\
0 & 0 & 0 & 0\\
\end{array}
\right),
\end{aligned}
\end{equation} 
\begin{equation}
\label{eq:app-irr-vertex-d-flex}
\begin{aligned}
\Gamma^{d}&= \frac{2 V_0 (V_0^2-4 V_1^2)( \chi_{\tau, \tau}^{(0),\text{PH}})^2+(3 V_0^2 +2 V_1^2 )\chi_{\tau, \tau}^{(0),\text{PH}} +V_0 }{( 1 - V_0 \chi_{\tau, \tau}^{(0),\text{PH}} )[(V_0 \chi_{\tau, \tau}^{(0),\text{PH}} +1 )^2-4 V_1^2 (\chi_{\tau, \tau}^{(0),\text{PH}})^2]} \left(
\begin{array}{cccc}
1 & 0 & 0 & 0 \\
 0 & 0 & 0 & 0 \\
 0 & 0& 0& 0 \\
0 & 0 & 0 &1 \\
\end{array}
\right) \\
&- \frac{V_1}{(V_0 \chi_{\tau, \tau}^{(0),\text{PH}} +1 )^2-4 V_1^2 (\chi_{\tau, \tau}^{(0),\text{PH}})^2} \left(
\begin{array}{cccc}
0& 0 & 0 & 0 \\
 0 & 1 & 0 & 0 \\
 0 & 0& 1& 0 \\
0 & 0 & 0 & 0\\
\end{array}
\right) +\left(2V_1 + \frac{2V_1^2 \chi_{\tau, -\tau}^{(0),\text{PH}}}{1- V_1 \chi_{\tau, -\tau}^{(0),\text{PH}}} \right)\left(
\begin{array}{cccc}
 0 & 0 & 0 & 1 \\
 0 & 0 & 0 & 0 \\
 0 & 0 & 0 & 0 \\
1 & 0 & 0 & 0 \\
\end{array}
\right)\\
&+\frac{V_1^2 \chi_{\tau,-\tau}^{(0),\text{PP}}}{V_1 \chi_{\tau,-\tau}^{(0),\text{PP}}-1} \left(
\begin{array}{cccc}
0& 0 & 0 & -2 \\
 0 & 1 & 0 & 0 \\
 0 & 0& 1& 0 \\
-2 & 0 & 0 & 0\\
\end{array}
\right) -\frac{V_0^2 \chi_{\tau,\tau}^{(0),\text{PP}}}{V_0 \chi_{\tau,\tau}^{(0),\text{PP}} - 1}\left(
\begin{array}{cccc}
1 & 0 & 0 & 0 \\
 0 & 0 & 0 & 0 \\
 0 & 0& 0 & 0 \\
0 & 0 & 0 & 1\\
\end{array}
\right).
\end{aligned} 
\end{equation} 

\end{widetext}

\bibliography{library-short}

\end{document}